\documentclass[iop]{emulateapj}

\usepackage{graphicx}
\usepackage{epstopdf}
\usepackage{amssymb}
\usepackage{hyperref}
\usepackage{mhchem}
\usepackage{amsmath}

\newcommand{\unit}[1]{\ensuremath{\, \mathrm{#1}}} 
\newcommand{\mol}[1]{\ensuremath{\mathrm{#1}}}

\newcommand{\molH}{\ensuremath{\mathrm{H_2}}}
\newcommand{\CO}{$\mol{CO}$}
\newcommand{\CI}{$\mol{[CI]}$}
\newcommand{\HCN}{$\mol{HCN}$}
\newcommand{\HCO}{$\mol{HCO^+}$}
\newcommand{\HNC}{$\mol{HNC}$}

\newcommand{\LNO}{$L_{\mathrm{HCN}}/L_{\mathrm{HCO}+}$}
\newcommand{\LOC}{$L_{\mathrm{HCO}+}/L_{\mathrm{CO}}$}
\newcommand{\LNC}{$L_{\mathrm{HCN}}/L_{\mathrm{CO}}$}
\newcommand{\LCN}{$L_{\mathrm{HNC}}/L_{\mathrm{HCN}}$}
\newcommand{\LCO}{$L_{\mathrm{HNC}}/L_{\mathrm{HCO}+}$}
\newcommand{\LOCO}{$L_{\mathrm{HCO}+}/L_{\mathrm{CO}}$}
\newcommand{\LNCO}{$L_{\mathrm{HCN}}/L_{\mathrm{CO}}$}

\begin{document}
\title{The dense gas in the largest molecular complexes of the
  Antennae: \HCN \ and \HCO \ observations of NGC 4038/39 using ALMA}
\author{Maximilien R.P. Schirm\altaffilmark{1},  Christine D. Wilson\altaffilmark{1}, Suzanne C. Madden\altaffilmark{2}, Dave L. Clements\altaffilmark{3}}
\altaffiltext{1}{Department of Physics and Astronomy, McMaster University, Hamilton, ON L8S 4M1 Canada; schirmmr@mcmaster.ca, wilson@physics.mcmaster.ca}
\altaffiltext{2}{Laboratoire AIM, CEA, Universit\'e Paris VII, IRFU/Service d'Astrophysique, Bat. 709, 91191 Gif-sur-Yvette, France}
\altaffiltext{3}{Astrophysics Group, Imperial College, Blackett Laboratory, Prince Consort Road, London SW7 2AZ, UK}

\begin{abstract}
We present observations of the dense molecular gas tracers \HCN, \HNC,
and \HCO \ in the $J=1-0$ transition using ALMA. 
We supplement our datasets with previous observations of \CO
\ $J=1-0$, which traces the total molecular gas content. We separate
the Antennae into 7  bright regions in which we detect emission from
all three molecules, including the nuclei of NGC 4038 and 
NGC 4039, 5 super giant molecular complexes in the overlap region
and 2 additional bright clouds. We find that the ratio
of \LNCO, which traces the dense molecular gas fraction, is greater in
the two nuclei (\LNCO$\sim 0.07 - 0.08$) than in the overlap region
(\LNCO$<0.05$). We attribute this to an increase in pressure due to
the stellar potential within the nuclei, similar to what has been seen
previously in the Milky Way and nearby spiral galaxies. Furthermore,
the ratio of \LCN$ \sim 0.3-0.4$ does not vary by more than a factor
of $1.5$ between regions. By comparing our measured ratios to PDR
models including mechanical heating, we find that the ratio of \LCN
\ is consistent with mechanical heating contributing $\gtrsim 5\% -
10\%$ of the PDR surface heating to the total heating budget.
Finally, the ratio of \LNO \ varies from $\sim 1$ in the nucleus of
NGC 4038 down to $\sim 0.5$ in the overlap region. The lower ratio in
the overlap region may be due to an increase in the cosmic ray rate
from the increased supernova rate within this region.  

\end{abstract}

\keywords{galaxies: individual (NGC 4038, NGC 4039) -  galaxies:
  interactions - galaxies: ISM - ISM: molecules}

\section{Introduction}

Merging and interacting galaxies play a fundamental role in the hierarchical evolution of galaxies (e.g. \citealt{steinmetz2002}). In a major merger, the turbulent motion generated by the gravitational interaction between the two merging galaxies can lead to a significant increase in the star formation rate, usually in the form of starbursts (e.g. see \citealt{hopkins2006} and references therein). For the most extreme mergers, this enhancement can culminate in an ultra luminous infrared galaxy (URLIG, $L_{IR} > 10^{12}$, \citealt{sanders1996}). In fact, it has been shown that most, if not all, ULIRGs are the direct result of an ongoing merger \citep{sanders1988, sanders1996, clements1996}. 

In the nearby universe, the closest example of a major merger is the Antennae (NGC 4038/39, Arp 244), with the two progenitor galaxies, NGC 4038 and NGC 4039, still in a relatively early merger stage. Within this system, the region where the two initial gas disks are believed to overlap has been dubbed the ``overlap region'' \citep{stanford1990}, while it has also been referred to as the ``interaction region'' \citep{schulz2007}. Young, massive ($> 10^6 M_\odot$), compact ``super star clusters'' (SSCs) are found throughout the overlap region \citep{whitmore1999,whitmore2010}. The overlap region also plays host to 5 super giant molecular complexes (SGMCs, \citealt{wilson2000}), massive ($> 10^8 M_\odot$) associations of molecular gas within which current and future star formation is expected. 

The molecular gas in the Antennae has been studied predominantly using
the molecular gas tracer \CO, which is excited via collisions with
\molH. The total molecular gas content within the Antennae, assuming a
Milky Way-like \CO-to-\molH \ conversion factor, is $\sim 2 \times
10^{10} \unit{M_{\odot}}$ \citep{gao2001}. Interferometric
observations of \CO \ $J=1-0$ show the presence of 100 super giant molecular clouds (SGMCs) throughout the system with a mass range $\sim 10^7 - 10^9 M_\odot$\citep{wilson2003}. More recently, \cite{schirm2014} analyzed observations of \CO \ $J=1-0$ to $J=8-7$ using a non-local thermodynamic equilibrium (non-LTE) radiative transfer analysis. They found that most of the molecular gas in the system is cool ($T_{kin} \sim 10 - 30 \unit{K}$) and intermediately dense ($n(\mol{H_2}) \sim 10^3 - 10^4 \unit{cm^{-3}}$), while a small fraction ($\sim 0.3 \%$) of the molecular gas is in a warm ($T_{kin} \gtrsim 100 \unit{K}$), dense ($n(\mol{H_2}) \gtrsim 10^4 \unit{cm^{-3}}$) phase. 

Recently, \cite{herrera2012} combined high-resolution Atacama Large Millimeter/submillimeter Array (ALMA) science verification observations of \CO \ $J=3-2$ in the Antennae with VLT/SINFONI imaging of the \molH \ $1-0 \unit{S(1)}$ transition, while \cite{whitmore2014} obtained ALMA cycle-0 observations of the overlap region in \CO \ $J=3-2$. By combining their data with observations from the Hubble Space Telescope (HST), and the Very Large Array (VLA), \cite{whitmore2014} identify regions within the overlap region corresponding to the various stages of star formation, from diffuse giant molecular clouds (GMCs) all the way to intermediate and old stellar clusters. Of particular interest is the very bright \CO \ $J=3-2$ emission found within SGMC 2, which \cite{whitmore2014} dubbed the ``firecracker'' and which is believed to be the precursor to a SSC. In addition, strong \molH \ $1-0 \unit{S(1)}$ emission is associated with the bright \CO \ $J=3-2$ emission \citep{herrera2012}. The mass and energetics of the cloud suggest a very high pressure ($P/k_B \gtrsim 10^8 \unit{K \ cm^{-3}}$),  while a lack of thermal radio emission indicates no star formation has yet occurred \citep{johnson2015}.

While \CO \ is the most commonly used tracer of molecular gas, due to its brightness and relatively high abundance in giant molecular clouds (GMCs), it may not be the ideal tracer for the star forming molecular gas. Dense gas tracers, such as \HCN \ \citep{gao2004a, gao2004b, papadopoulos2007}, \HNC \ \citep{talbi1996}, and \HCO \  \citep{graciacarpio2006}, can be used to study the molecular gas most directly linked to star formation. This relation culminates in the form of a tight relationship between the luminosity of \HCN \ ($L_{\mol{HCN}}$) and the infrared (IR) luminosity ($L_{IR}$), tighter than the correlation between \CO \ and the infrared luminosity \citep{liu2010}. In fact, this relationship holds in the form of a constant ratio of $L_{\mol{HCN}}/L_{IR}$ from individual giant molecular clouds (GMCs) all the way up to luminous infrared galaxies (LIRGs) \citep{wu2005}.  Recently, \cite{bigiel2015} used CARMA data to compare HCN and other dense tracers to the far-infrared luminosity to trace star formation efficiency and dense gas fraction. The critical densities of \HCN, \HNC, and \HCO \ are all $n_{cr} > 10^5 \unit{cm^{-3}}$ for the $J=1-0$ transition \citep{loenen2007}, significantly higher than \CO \ $J=1-0$ ($n_{cr} \sim 3 \times 10^3 \unit{cm^{-3}}$, \citealt{loenen2007}). 

Because \HCN, \HNC, and \HCO \ in the $J=1-0$ transition all have
similar critical densities, the relative strengths of the lines of
these molecules are driven largely by ongoing physical
processes. Infrared (IR) pumping can enhance \HCN \ emission in a
system with a strong background mid-IR field \citep{sakamoto2010},
while \HCO \ can be destroyed via dissociative recombination in the
presence of a significant abundance of electrons. \HCN \ and \HNC
\ are isotopomers, and their relative abundance is driven by the
temperature of the dense molecular gas \citep{talbi1996}. Line ratios
of these molecules can be used as a diagnostic for cosmic rays
\citep{meijerink2011}, photon dominated regions (PDRs,
\citealt{kazandjian2012}), and mechanical heating \citep{loenen2008,
  kazandjian2012}.  Thus, these lines can be used to study the
distribution of dense molecular gas within a system and 
to constrain the relative fraction of dense gas from
region to region as well as the relative
importance of physical processes such as mechanical heating.

In this paper, we present observations of the $J=1-0$ transition of
\HCN, \HNC, and \HCO \ in the Antennae using ALMA. We adopt a distance
to NGC 4038/39 of $D = 22 \unit{Mpc}$ \citep{schweizer2008},
corresponding to an angular scale of $107 \unit{pc/''}$. 
We combine these observations with previous observations of \CO \ $J=1-0$ from \cite{wilson2000, wilson2003}. In this work, we detail the observations and data reduction in Section \ref{obsSect}, while we discuss separating the emission into the brightest regions in Section \ref{brightSect}. We present various line ratios of \HCN, \HNC, \HCO, and \CO \ in Section \ref{lineRatSect}, and we discuss the implications of the measured line ratios in Section \ref{discSect}. 



\section{Observations} \label{obsSect}

NGC 4038 was observed in two pointings over the course of 2 nights each in March, 2013 and April, 2014 as part of ALMA Cycle-1. One pointing was positioned to cover the nucleus of NGC 4038 and the western loop ($12\unit{h}01\unit{m}51.7763\unit{s}, -18^{\circ}52'02.892''$), while the other was positioned to cover both the overlap region and the nucleus of NGC 4039 ($12\unit{h}01\unit{m}54.5945\unit{s}, -18^{\circ}52'50.891''$), with a total on-source integration time of $20356 \unit{seconds}$. The antenna array configuration was C32-4. The bandpass calibrators used were J1256-0547, J1246-2547, and J1130-1449, while J1215-1731 was used as the phase calibrator and Titan was used as the amplitude calibrator throughout. 

Band 3 was used, with one spectral window set to observe both \HCN \ and \HCO \ simultaneously. A second spectral window was aligned to search for $\mol{HNC}$ \ emission throughout the Antennae. The total usable bandwidth for each of these spectral windows was $1875 \unit{MHz}$ with a channel spacing of $0.488 \unit{MHz}$. The two remaining spectral windows were centred at $101.0 \unit{GHz}$ and $102.875 \unit{GHz}$ to observe the continuum emission across the Antennae. The bandwidth of the two continuum windows was $2000\unit{MHz}$ with a channel spacing of $15.625 \unit{MHz}$. 

We reduce these data using the Common Astronomy Software Applications package version 4.2. We start by running the script provided to the PI as part of the data delivery in order to calibrate the \emph{uv}-datasets before splitting out the target from the calibrated dataset. We further split out the spectral window which contains the \HCN \ and \HCO \ lines, and the spectral window which contains the \HNC \ line. We subtract the background continuum emission by fitting a first order polynomial to the line-free channels of each spectral window using the \verb+uvcontsub+ task. 

We create dirty data-cubes for each of the \HCN, \HCO, and \HNC \ $J=1-0$ transitions with a channel width of $5 \unit{km \ s^{-1}}$ and cell size of $0.3''$. The restoring beams of the cleaned data cubes when using Briggs weighting with robust $= 0.5$ are $1.85'' \times 1.51'', -79.2^\circ$ for the \HCO \ line and $1.86'' \times 1.52'', -79.8^\circ$ for the \HCN \ line (the \HNC \ line was too weak to detect detect at the native ALMA resolution). We set the restoring beam for the \HCN \ and \HCO \ transitions to the \HCN \ beam and create clean data-cubes by placing clean boxes around $>5\sigma$ emission in the dirty data-cubes, being careful not to select artifacts. We clean down to a threshold of $1.1 \unit{mJy \ beam^{-1}}$, corresponding to $2\sigma$ in both dirty data cubes as measured in the line-free channels ($\sigma = 0.55 \unit{mJy \ beam^{-1}}$).

In addition, we create dirty and clean data cubes to match the beam of the \CO \ $J=1-0$ data cube from \cite{wilson2000, wilson2003}. We set the \emph{uv}-taper on the outer baselines to $5.0''\times 2.0'', 1.45^\circ$ for \HCN, \HCO, and \HNC \ with a restoring beam of $4.91'' \times 3.15'', 1.45^\circ$. We use the same channel width and cell size as before, and draw clean boxes around $5\sigma$ emission. We clean down to a threshold of $1.6 \unit{mJy \ beam^{-1}}$, corresponding to $2\sigma$ of the tapered dirty data cubes ($\sigma = 0.8 \unit{mJy \ beam^{-1}}$).

\begin{figure*}[ht] 
\centering
$\begin{array}{cc}
	\includegraphics[width=0.5\linewidth]{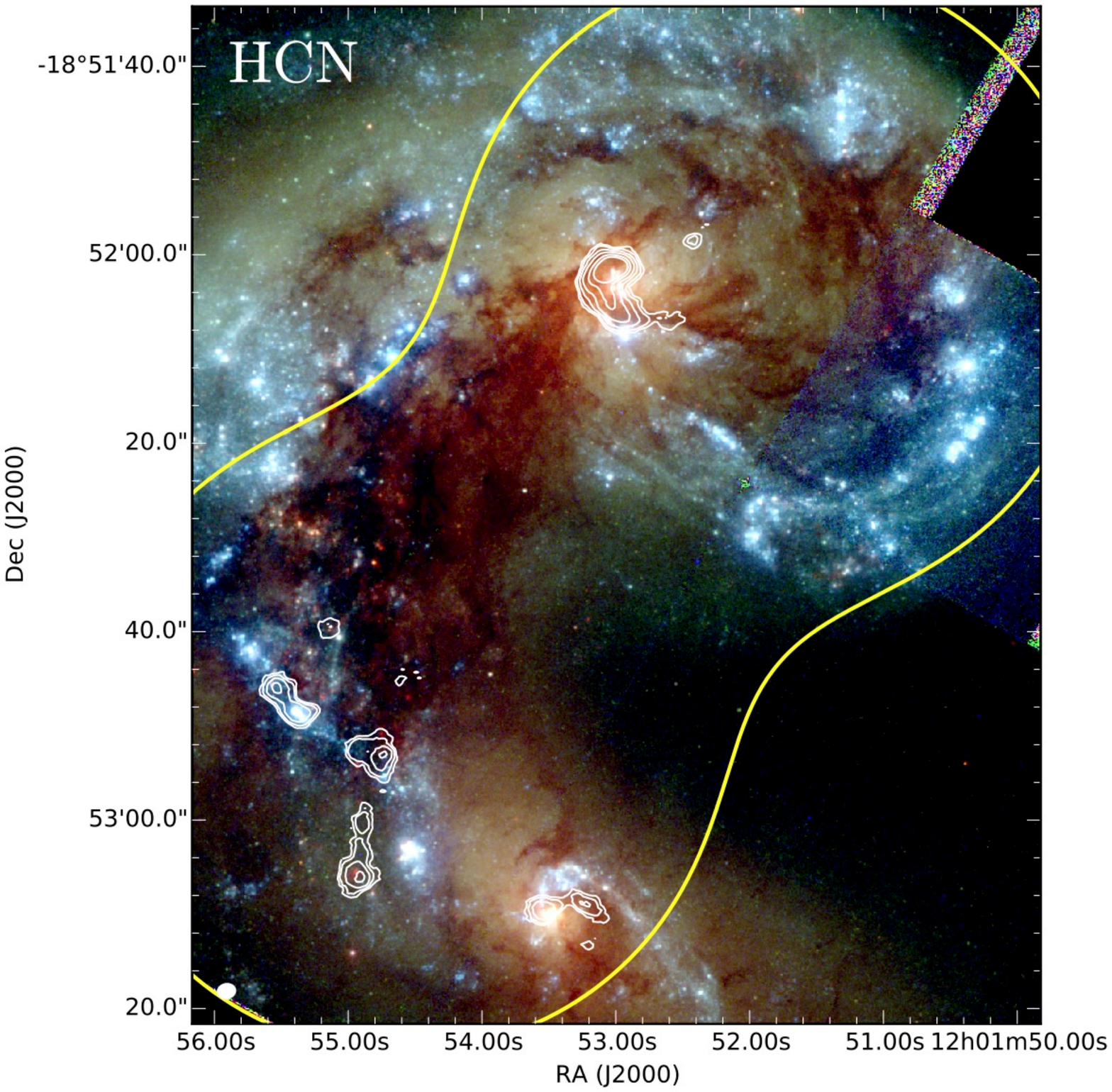}   &
	\includegraphics[width=0.5\linewidth]{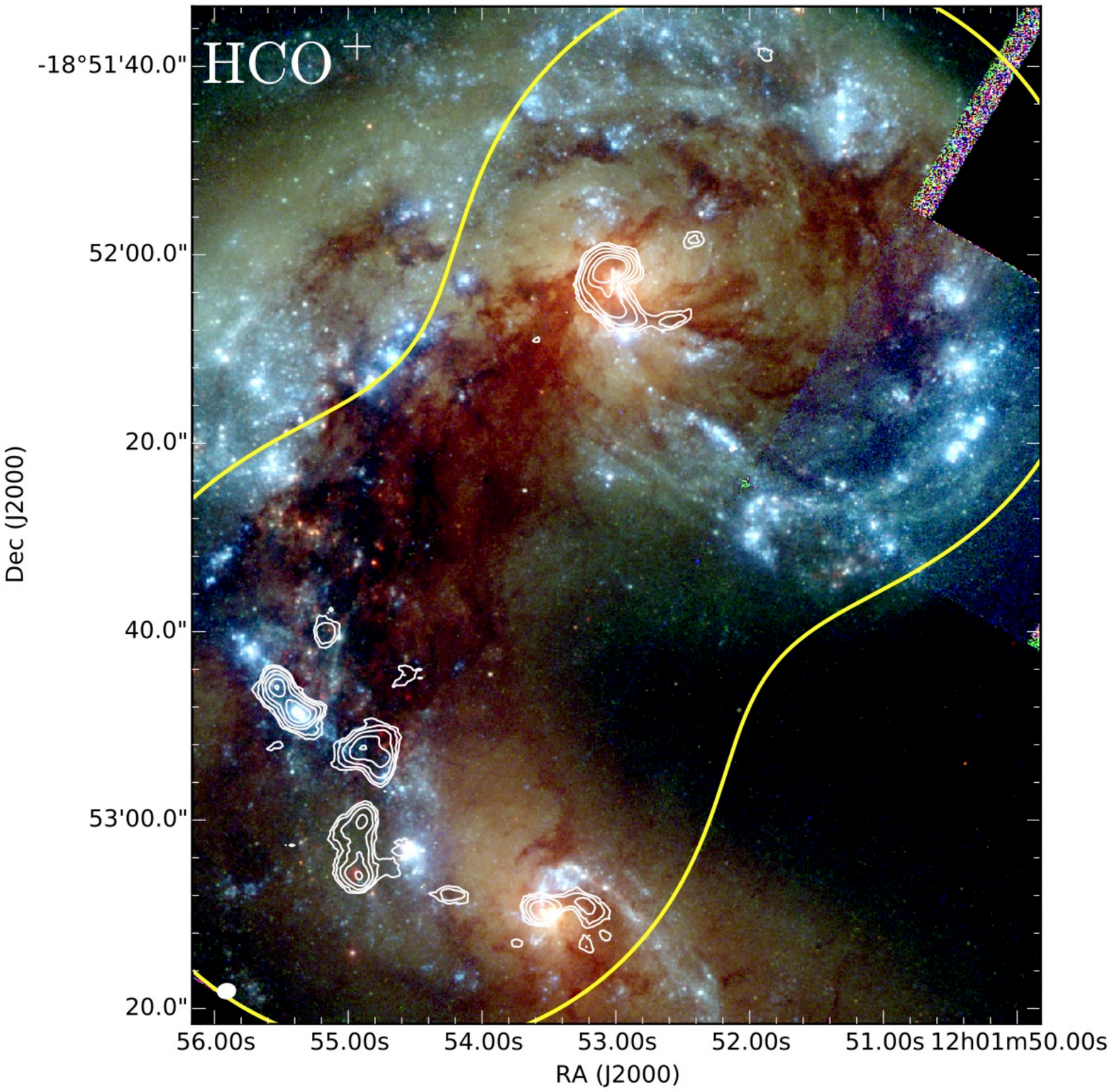}  
\end{array}$
\caption[]{\HCN \ (\emph{left}) and \HCO \ (\emph{right}) moment-0
  contours overlaid on the same HST image of the Antennae. The
  contours shown have not been corrected for the primary beam. The HST
  image was created using data from \cite{whitmore1999}. The \HCN
  \ and \HCO \ contours, shown in white, correspond to $(3, 5, 9, 15,
  25, 35) \times (2.8\times 10^{-2} \unit{Jy \ beam^{-1} \ km
    \ s^{-1}})$. The beam of the dense gas tracer observations is
  shown by a white ellipse in the lower left corner. The combined
  half-power beam width of the two pointings is shown by the yellow contour.}
\label{denseOnHST}
\end{figure*}

\begin{figure*}[ht] 
\centering
$\begin{array}{ccc}
	\includegraphics[width=0.33\linewidth]{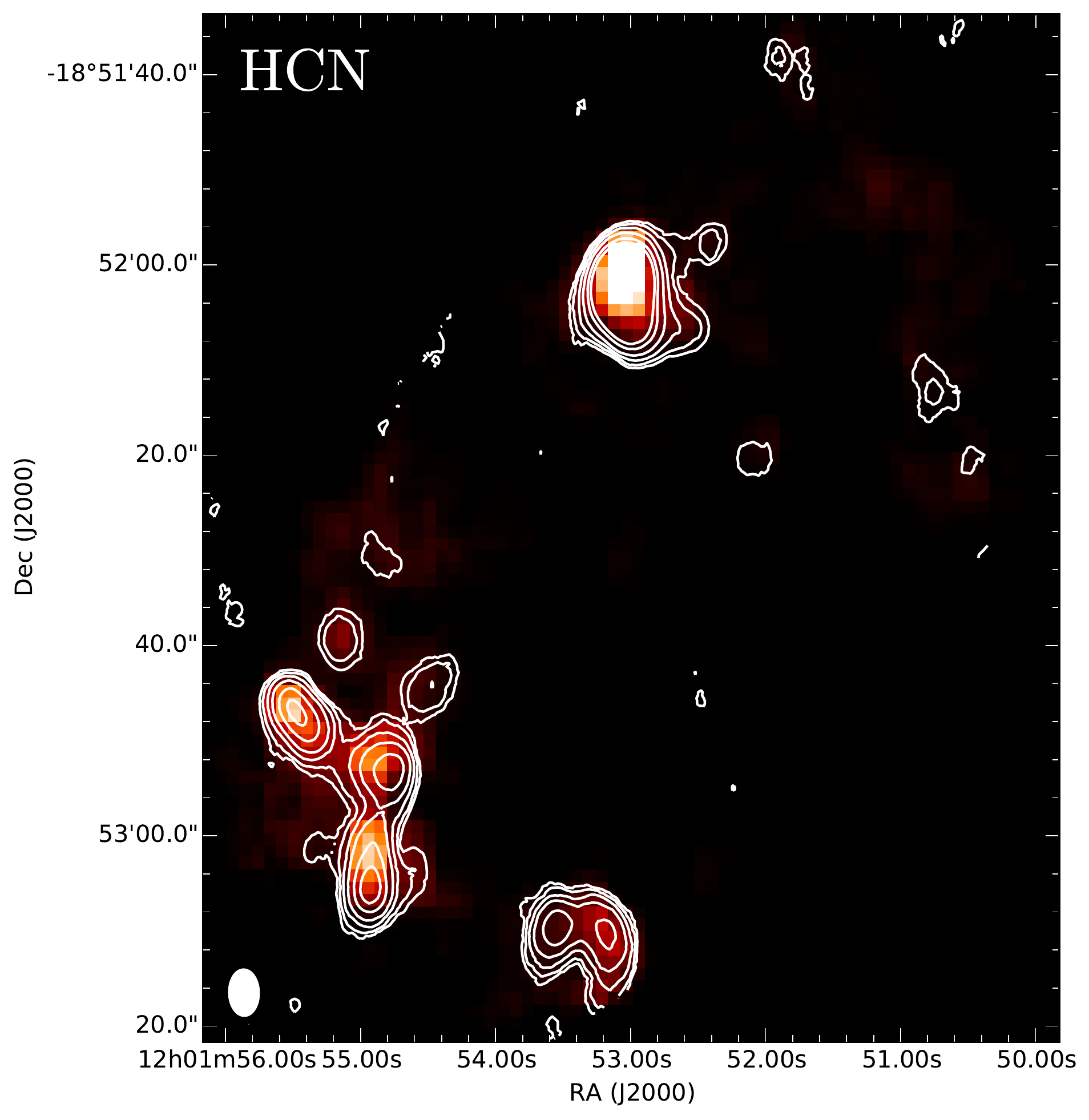}   &
	\includegraphics[width=0.33\linewidth]{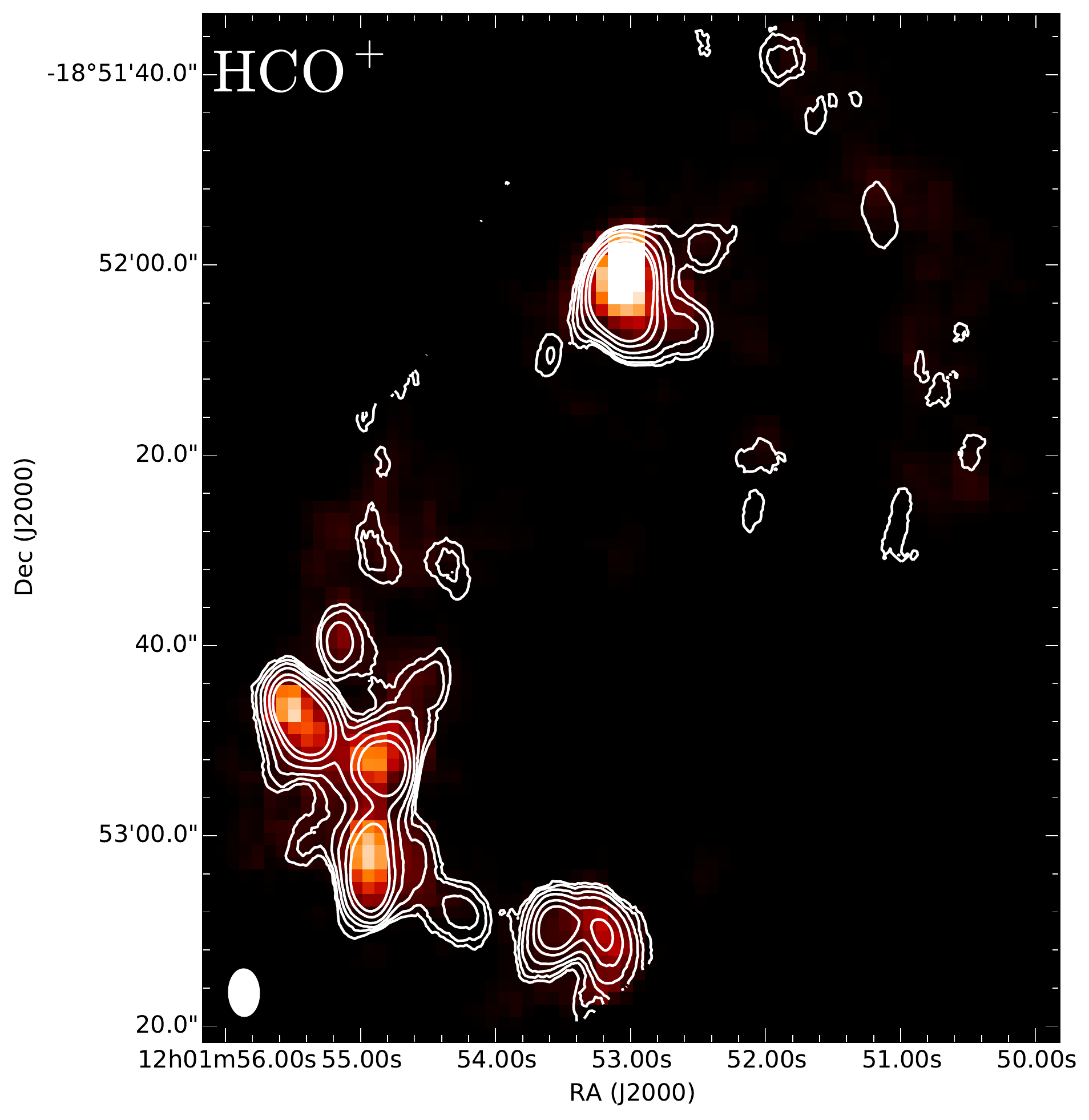}  &
	\includegraphics[width=0.33\linewidth]{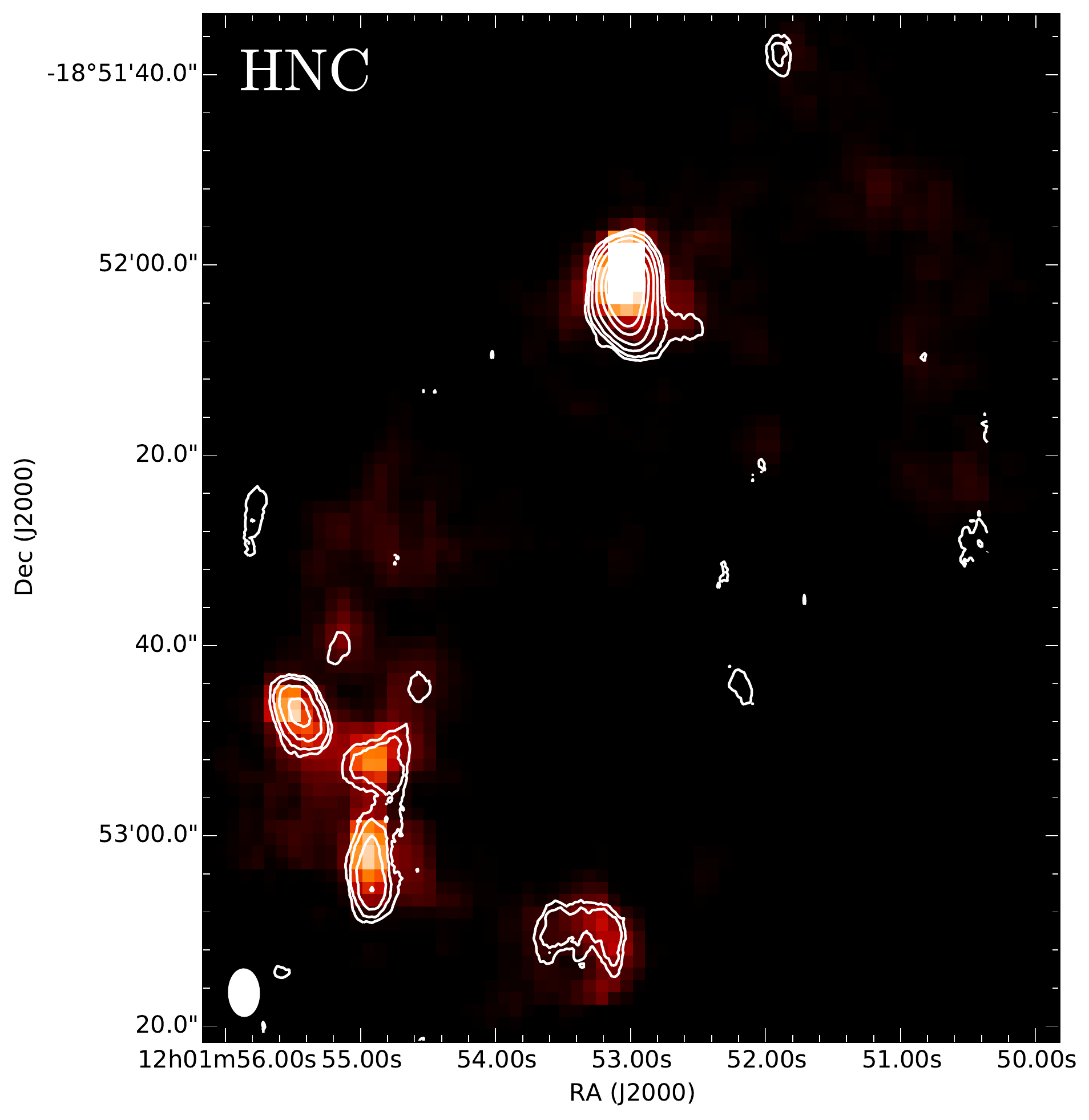} 
	\end{array}$
\caption[]{\HCN \ (\emph{left}), \HCO \ (\emph{middle}), and \HNC
  \ (\emph{right}) moment-0 contours beam-matched to the \CO \ $J=1-0$
  observations from \cite{wilson2003} overlaid on the same \CO
  \ $J=1-0$ moment-0 map. The contours correspond to $(3, 5, 9, 15,
  25, 35) \times (3.6 \times 10^{-2} \unit{Jy \ beam^{-1} \ km
    \ s^{-1}})$. The beam is shown by the white ellipse in the lower
  left corner of each image. These contours have been corrected for the
  fall-off in sensitivity due to the primary beam.}
\label{denseOnCO}
\end{figure*}

We create moment maps including emission above $2\sigma$. The un-tapered moment 0 maps are shown in Figure \ref{denseOnHST}, while the tapered moment 0 maps are shown in Figure \ref{denseOnCO}, along with the \CO \ $J=1-0$ moment 0 map from \cite{wilson2000, wilson2003}. Finally, we correct all of our \HCN, \HCO, and \HNC \ moment-0 maps for the primary beam. The largest correction occurs in the nucleus of NGC 4039, where the primary beam correction is on the order of $\sim 0.6-0.7$ for both the native ALMA resolution maps and the \ce{CO} $J=1-0$ resolution maps. The \ce{CO} $J=1-0$ map from \cite{wilson2003} has already been corrected for the primary beam. 

\cite{gao2001} observed the Antennae in \HCN \ $J=1-0$ in two pointings: one centered on the nucleus of NGC 4038, and the other on the brightest \CO \ emission from the overlap region. The full-width half-maximum of their beam was $\sim 72''$ and their two pointings include both nuclei and the overlap region. They calculate a total \HCN \ $J=1-0$ luminosity of $0.8 \times 10^{8} \unit{K \ km \ s^{-1} \ pc^{2}}$ (adjusted to our adopted distance of $22 \unit{Mpc}$), while we measure a total of \HCN \ $J=1-0$ luminosity of  $\sim 0.7 \times 10^{8} \unit{K \ km \ s^{-1} \ pc^{2}}$ in our ALMA maps. Our detected \HCN \ emission is well within their two beams, and we do not expect significant \HCN \ emission outside of our observed region. This comparison suggests that we recover $\sim 90\%$ of the total \HCN \ emission in our ALMA observations.

We estimate the measurement uncertainties in our moment maps by
first determining the number of channels included in each pixel of the
moment map ($N_{chan}$). We calculate the uncertainty in each pixel as
$\sqrt{N_{chan}} \sigma \Delta v$, where $\sigma$ is the root-mean
squared uncertainty measured from our line-free channels of our data
cubes, and $\Delta v = 5 \unit{km/s}$ is the channel
width. Furthermore, when comparing our dense gas tracers to our \CO
\ $J=1-0$ map, we add a $5\%$ calibration uncertainty\footnote{ALMA
  Cycle 1 Technical Handbook. Available at
  \url{https://almascience.nrao.edu/documents-and-tools/cycle-1/alma-technical-handbook}}. The
calibration uncertainty of the \CO \ $J=1-0$ map is $20 \%$
\citep{wilson2003}. While individual luminosities can vary
  depending on the size of the aperture chosen, line ratio measured in
the same apertures from maps with the same beam size should be quite
accurate measures of the average line ratio over a given region. Line
ratios involving the \HNC \ line may be somewhat more uncertain as
this is the weakest line and so the 2$\sigma$ clipping used to make
the moment maps is a larger fraction of the peak flux.

\section{The brightest regions in the Antennae} \label{brightSect}

In this paper, we will focus on the \HCN, \HNC, and \HCO \ emission from the brightest regions in the Antennae. A full catalogue of clouds in both \HCN, \HNC, and \HCO \ will be published in a future paper, along with a cloud-by-cloud analysis of the emission. 

\cite{wilson2000} identified and isolated the 7 brightest regions in their \CO \ $J=1-0$ map of the Antennae using the clump identification algorithm CLFIND \citep{williams1994}. These regions include both nuclei (NGC 4038 and NGC 4039), and 5 large clouds in the overlap region, dubbed Super Giant Molecular Complexes (SGMCs).  In addition to these 7 regions, we identify two additional bright regions, C6 and C7. These clouds correspond to clouds 16, 17 and 18 (C6), and clouds 67 and 74 (C7) in the \CO \ cloud catalogue published in \cite{wilson2003}. 

\begin{figure}[ht] 
	\centering
	\includegraphics[width = \linewidth]{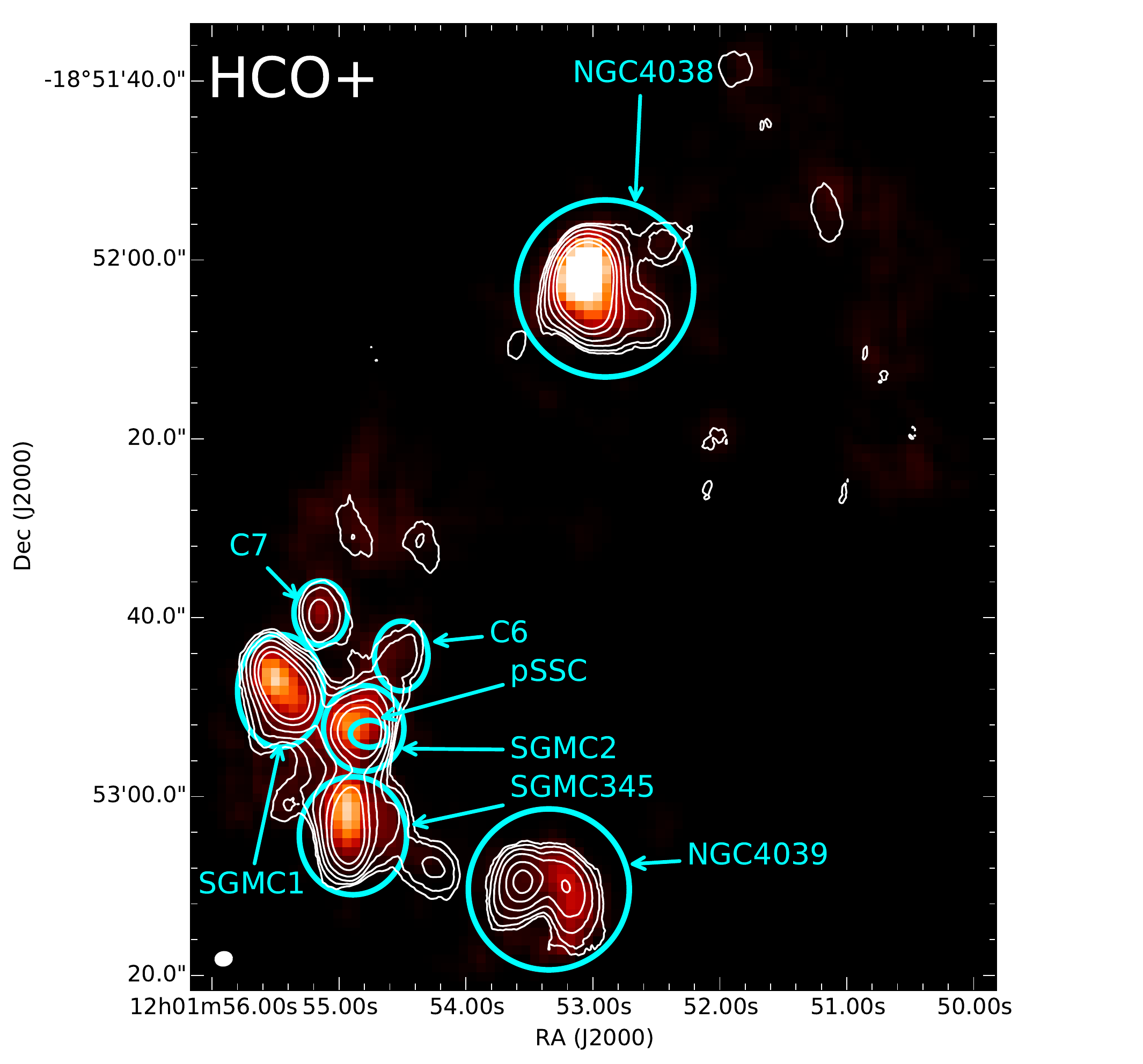}
	\caption[]{Map of the brightest regions in the Antennae. The background image is the \CO \ $J=1-0$ map from \cite{wilson2000} while the \HCO \ $J=1-0$ contours from Figure \ref{denseOnCO} are overlaid. The cyan ellipses indicate the location of the two nuclei, the 5 SGMCs and the two additional clouds, C6 and C7, along with the pre-super star cluster ``pSSC''.}
	\label{regionsHCOonCO}
\end{figure}

We place elliptical apertures around each of the 9 brightest regions to measure the total \HCN, \HNC, \HCO, and \CO \ fluxes ($S_{mol}$, Table \ref{fluxTab}) in the moment-0 map for the $J=1-0$ transition for each molecule (Figure \ref{regionsHCOonCO}). The total fluxes and total luminosities of the nuclei, SGMCs and clouds C6 and C7 are measured at the \CO \ $J=1-0$ resolution for all 4 molecules.

For the 5 SGMC regions, we are unable to distinguish between SGMC 3, 4
and 5 in the moment-0 map and so we combine these three SGMCs into
SGMC 3+4+5. We can, however, distinguish between them in
velocity-space, as shown in \cite{wilson2000} (see their Figure
  5). Using the measured local standard of rest velocity 
($V_{lsr}$) along with the velocity widths ($\Delta V_{FWHM}$) from
\cite{wilson2000}, we create moment-0 maps for the velocity range
spanned by each SGMC. We measure the flux and luminosity for each
cloud in their own moment-0 map using the same aperture as used for
SGMC 3+4+5. There are 2 channels of overlap in the velocity ranges for
SGMC 4 and 5 as measured by \cite{wilson2000}; we separate the
two SGMCs by including one channel in each of the moment maps of 
  SGMC 4 and 5.  

\cite{johnson2015} identified a small region in SGMC 2 in their ALMA
\CO \ $J=3-2$ map of the overlap region which they believe to be the
precursor to a super star cluster, which we have dubbed ``pre-SSC'' in
this work. The resolution of their \CO \ $J=3-2$ map is considerably
better than that of our observations (beam size $= 0.56'' \times
0.43''$), while the size of pre-SSC as measured using the CPROPS
program \citep{rosolowsky2006} is $\sim 0.66'' \times 0.55''$, less
than the size of the beam of our \HCN \ and \HCO \ observations. We
created a moment-0 map using the velocity range measured
for the pre-SSC
by \cite{johnson2015} and our data cubes at the native ALMA
resolution. However, even in this map, the pre-SSC is not visible as a
separate source. The \HCN \ and \HCO \ peak
fluxes measured at the location of the pre-SSC are 0.28 and 0.45 Jy
beam$^{-1}$ 
km s$^{-1}$, respectively; however, these fluxes almost certainly
contain significant contamination from SGMC 2 and so should be
regarded as upper limits to the true flux of the pre-SSC.

\begin{deluxetable*}{lcccccc} 
\tablecaption{Measured flux densities of the molecular lines \label{fluxTab}} 
\tablehead{ &\colhead{$S_{\mathrm{HCN}}$\tablenotemark{a}} &\colhead{$S_{\mathrm{HCO}+}$\tablenotemark{a}} &\colhead{$S_{\mathrm{HNC}}$\tablenotemark{a}} &\colhead{$S_{\mathrm{CO}}$\tablenotemark{b}} \\ 
\colhead{Region ID} & $(\mol{Jy \ km \ s^{-1}})$ &$(\mol{Jy \ km \ s^{-1}})$ &$(\mol{Jy \ km \ s^{-1}})$ &$(\mol{Jy \ km \ s^{-1}})$ & 
}\startdata 
NGC 4038& $14.32 \pm 0.08$& $14.23 \pm 0.08$& $5.58 \pm 0.06$& $290 \pm 5$
\\NGC 4039& $4.03 \pm 0.08$& $6.51 \pm 0.09$& $1.06 \pm 0.05$& $100 \pm 5$
\\SGMC 1& $2.59 \pm 0.05$& $6.63 \pm 0.06$& $1.01 \pm 0.04$& $117 \pm 2$
\\SGMC 2& $2.12 \pm 0.04$& $4.50 \pm 0.05$& $0.59 \pm 0.03$& $92 \pm 2$
\\SGMC 3& $0.69 \pm 0.03$& $1.82 \pm 0.03$& $0.25 \pm 0.02$& $61 \pm 2$
\\SGMC 4& $1.32 \pm 0.03$& $3.03 \pm 0.04$& $0.46 \pm 0.02$& $67 \pm 3$
\\SGMC 5& $0.94 \pm 0.03$& $1.35 \pm 0.03$& $0.30 \pm 0.02$& $30 \pm 2$
\\SGMC 3+4+5& $3.34 \pm 0.06$& $6.18 \pm 0.06$& $1.10 \pm 0.04$& $128 \pm 3$
\\C6& $0.40 \pm 0.02$& $0.45 \pm 0.02$& $0.14 \pm 0.01$& $15 \pm 1$
\\C7& $0.34 \pm 0.02$& $0.53 \pm 0.02$& $0.13 \pm 0.01$& $22.5 \pm 0.8$
\enddata 
\tablecomments{The reported uncertainties are the measurement uncertainties. \\ 
To convert into luminosity with units of $\mol{K \ km \ s^{-1} \ pc^2}$, multiply fluxes by \\ 
 $1.5666 \times 10^{10}(\nu_0/\mol{GHz})^{-2}$. See text for additional details.
\tablenotetext{a}{The calibration uncertainty for ALMA band 3 in Cycle 1 is $5\%$ (see text)}  
\tablenotetext{b}{The calibration uncertainty for the \ce{CO} observations is $20 \%$ (see text and \citealt{wilson2003})} 
} 
\end{deluxetable*}

\section{Line Ratios}\label{lineRatSect}

\begin{figure*}[htp!] 
\centering
$\begin{array}{ccc}
	\includegraphics[height=0.37\linewidth]{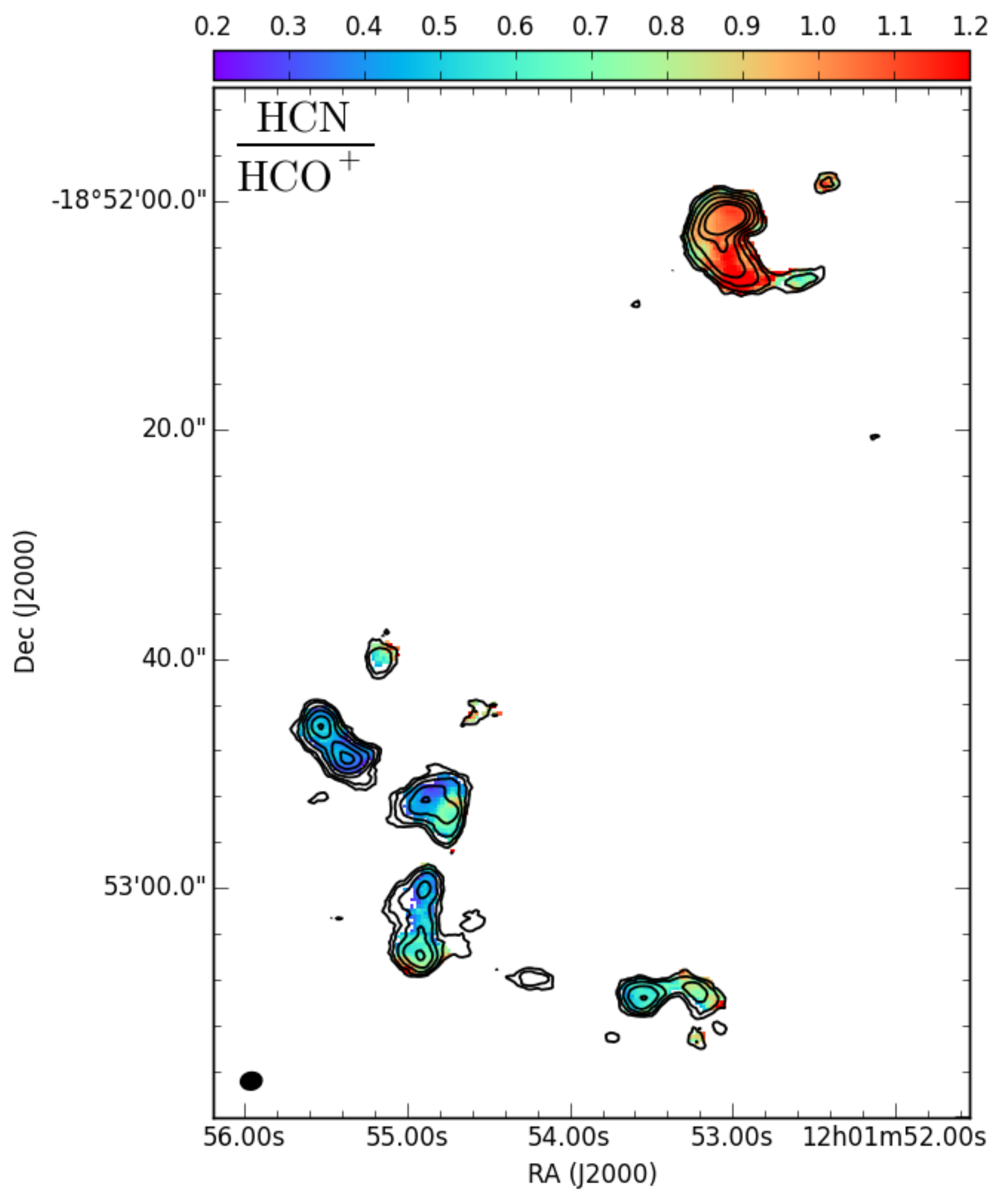}   &
	\includegraphics[height=0.37\linewidth]{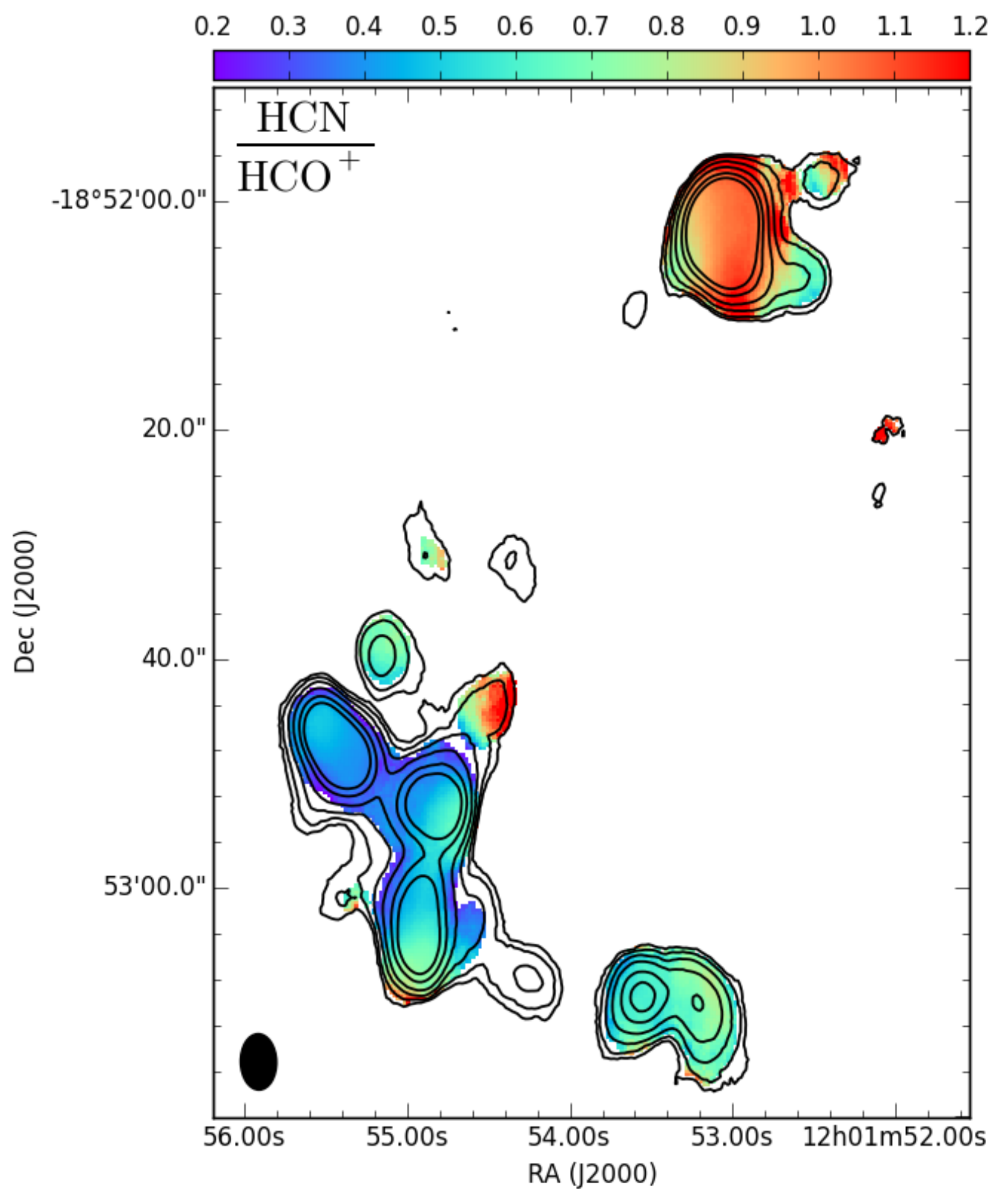} &
	\includegraphics[height=0.37\linewidth]{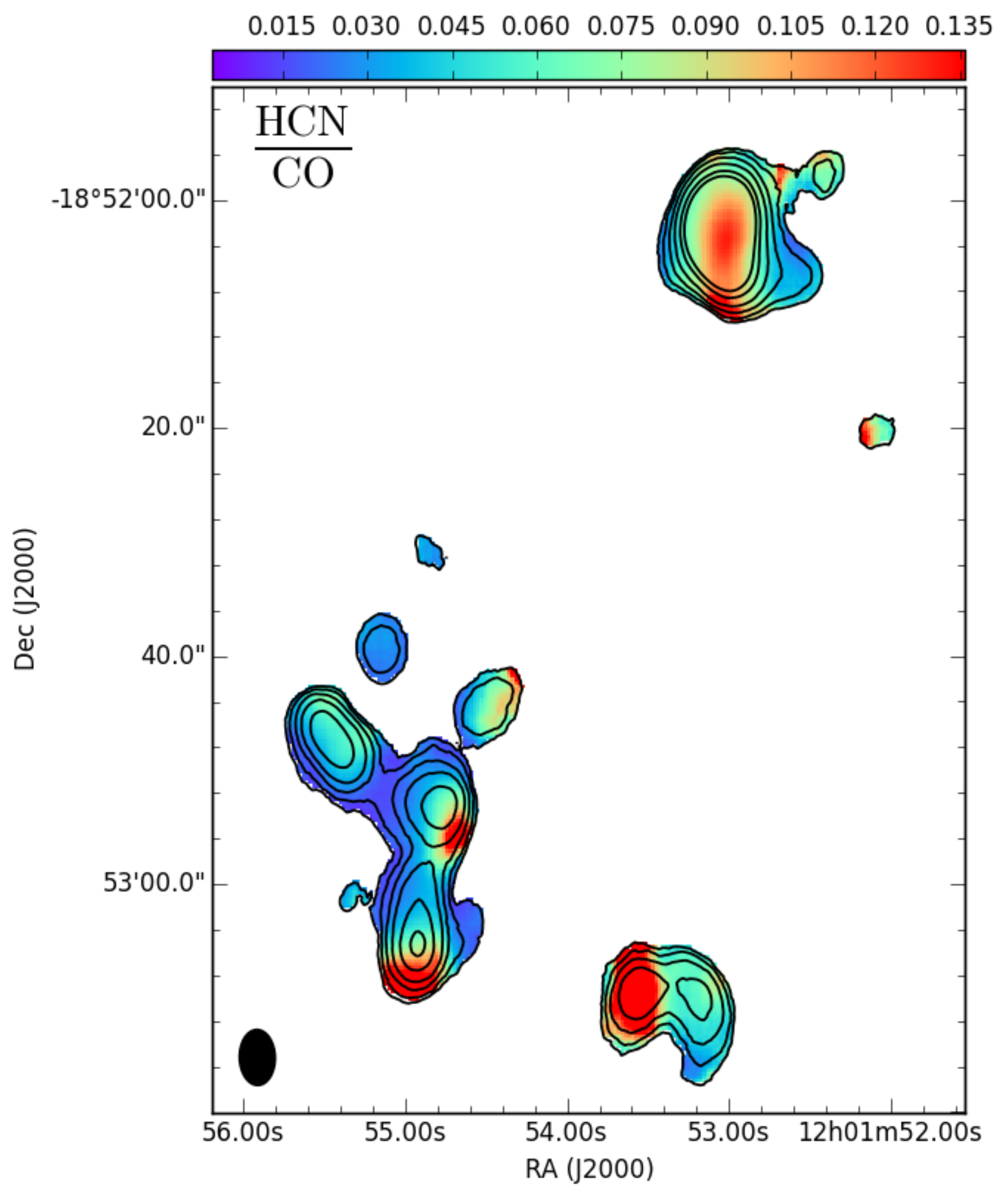}
\end{array}$
\caption[]{\HCN/\HCO \ intensity ratio maps at the native ALMA beam
  (\emph{left}) and \CO \ $J=1-0$ beam (\emph{middle}), along with the
  \HCN/\CO \ intensity ratio map (\emph{right}). The \HCO \ contours from
  Figure \ref{denseOnHST} (\emph{left}) and Figure \ref{denseOnCO}
  (\emph{middle}) are overlaid on the \emph{left} and \emph{middle}
  line ratio maps, while the \HCN \ contours from Figure
  \ref{denseOnCO} (\emph{left}) are overlaid on the \emph{right} line
  ratio map. 
The nucleus of NGC 4038 exhibits the highest ratio of \HCN/\HCO ($\sim 1$) while the same line ratio in the overlap region is about a factor of $2$ smaller.}
\label{HCNHCOratioMap}
\end{figure*}

We calculate the \HCN/\HCO \ and \HCN/\CO \ line ratio maps for both
the native resolution and \CO \ beam matched moment-0 maps by dividing
the moment-0 maps directly (Figure \ref{HCNHCOratioMap}). The
difference between the primary beam correction for the \HCN \ and \HCO
\ moment-0 maps is minimal, and so we use the non-primary beam
corrected maps for this line ratio. For the line ratio of \HCN/\CO, we
use the primary beam-corrected maps. In addition, we include only
pixels with a $3\sigma$ detection of both molecular transitions 
  in the moment-0 maps. 

\subsection{Ratios of the dense gas tracers}

\begin{figure}[ht!] 
	\centering
	\includegraphics[width = \linewidth]{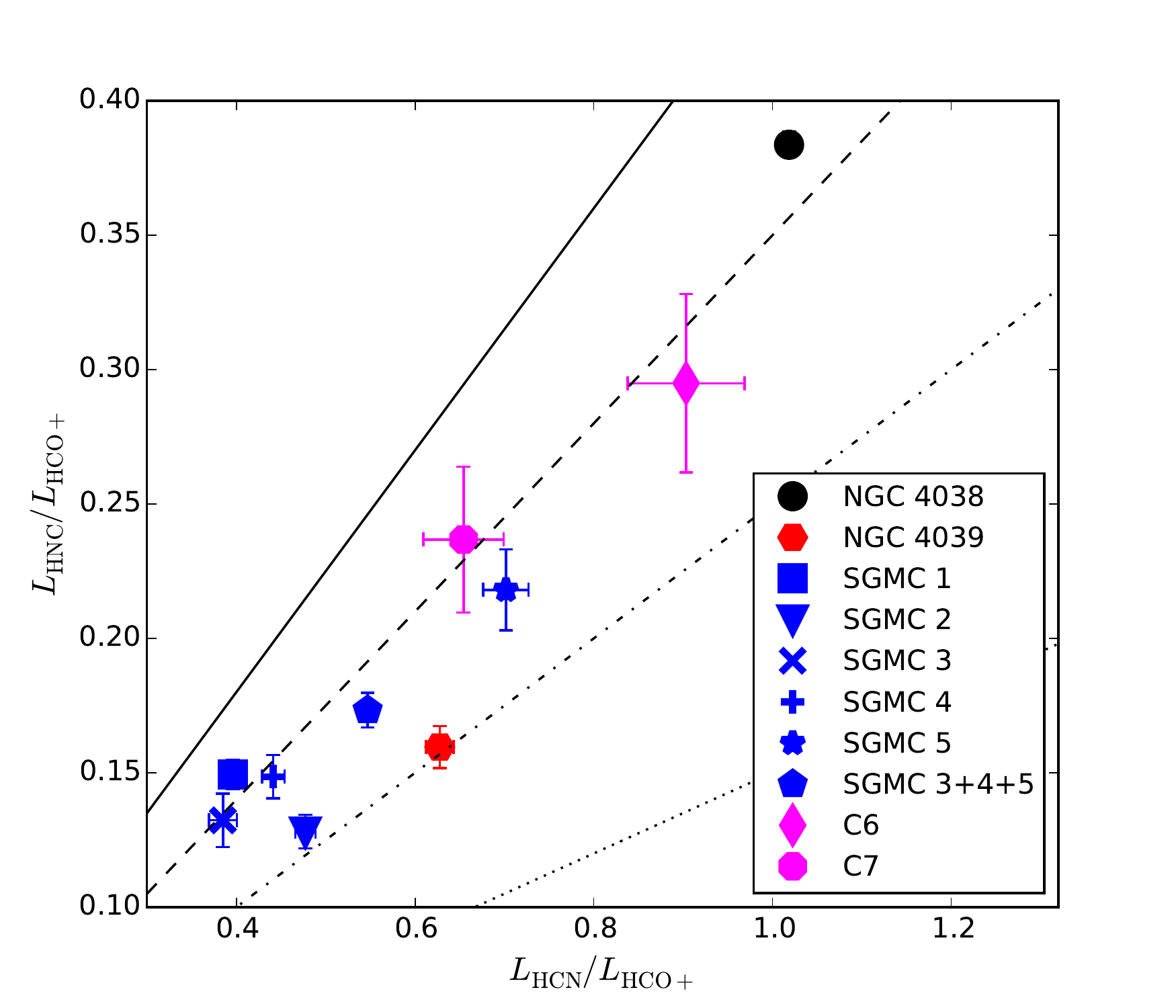}
	\caption[]{Luminosity ratios of the dense gas tracers for the brightest regions in the Antennae calculated at the \CO \ $J=1-0$ beam. The error bars correspond to measurement uncertainties in the line ratios. The lines correspond to lines of constant \LCN, with \LCN $= 0.15$ (dotted line), $0.25$ (dash-dot line), $0.35$ (dashed line) and $0.45$ (solid line). The region SGMC 3+4+5 corresponds to the region in the moment-0 map spanned by SGMC 3, 4 and 5, while the values for SGMC 3, 4 and 5 were calculated from moment maps created based on the velocity range calculated in \cite{wilson2000}}
	\label{denseRatPlot}
\end{figure}

We calculate the ratios of the luminosities of the dense gas tracers \HCN, \HNC, and \HCO \ for the brightest regions in the Antennae (see Section \ref{brightSect}). We use the fluxes we measure in Table \ref{fluxTab} at the OVRO beam size, and convert them to luminosities using the following formula \citep{wilson2008} 

\begin{multline}
    \frac{L'}{\mol{K \ km \ s^{-1} \ pc^2}} = 3.2546 \times 10^7 \left(\frac{S_{\mol{mol}}}{\mol{Jy \ km \ s^{-1}}} \right) \left(\frac{D_L}{\mol{Mpc}}\right)^{2} \\
    \times \left(\frac{\nu_0}{\mol{GHz}} \right)^{-2} (1+z)^{-1}
\end{multline}

\noindent where $S_{\mol{mol}}$ is the flux, $D_L=22\unit{Mpc}$ is the distance, $\nu_0$ is the rest frequency of the transition, and $z = 0.005477$ is the redshift. We show these ratios of \LCN, \LCO, and \LNO in Table \ref{denseRatioTable} and we plot the ratios of \LCO \ and \LNO \ in Figure \ref{denseRatPlot}. 

With the exception of C6, the nucleus of NGC 4038 shows values of \LNO
\ and \LCO \ more than $1.5$ times greater than the other regions. For
C6, the ratios of \LNO \ and \LCO \ are only $\sim 1.1$ and $\sim 1.3$
times less than in NGC 4038.  Of the remaining regions, SGMCs 1, 2, 3
and 4 exhibit similar values for both the ratio of \LNO \ ($\sim 0.39
- 0.48$) and the ratio of \LCO \ ($\sim 0.13-0.15$), suggesting
similar 
dense gas properties across the 4 regions. The nucleus of NGC 4039
exhibits slightly larger values for both (\LNO $\sim 0.63$, \LCO$\sim
0.16$). Of the SGMCs, SGMC 5 exhibits the largest value for both line
ratios (\LNO $\sim 0.7$, \LCO$\sim 0.22$); this enhanced line ratio
can also be seen in the high-velocity end of the spectrum for this
region in \citet{bigiel2015}.

The value of \LCN \ varies by less than a factor of 1.5 across all 9 regions. The largest deviations occur for the nucleus of NGC 4039 and SGMC 2, with \LCN $=0.25$ and \LCN $=0.27$ respectively. Of the remaining 8 regions, \LCN \ is nearly constant, varying by only $\sim 20\%$ (\LCN $\sim 0.31 - 0.38$). 

\subsection{Ratios to \CO}

\begin{figure}[ht!] 
	\centering
	\includegraphics[width = \linewidth]{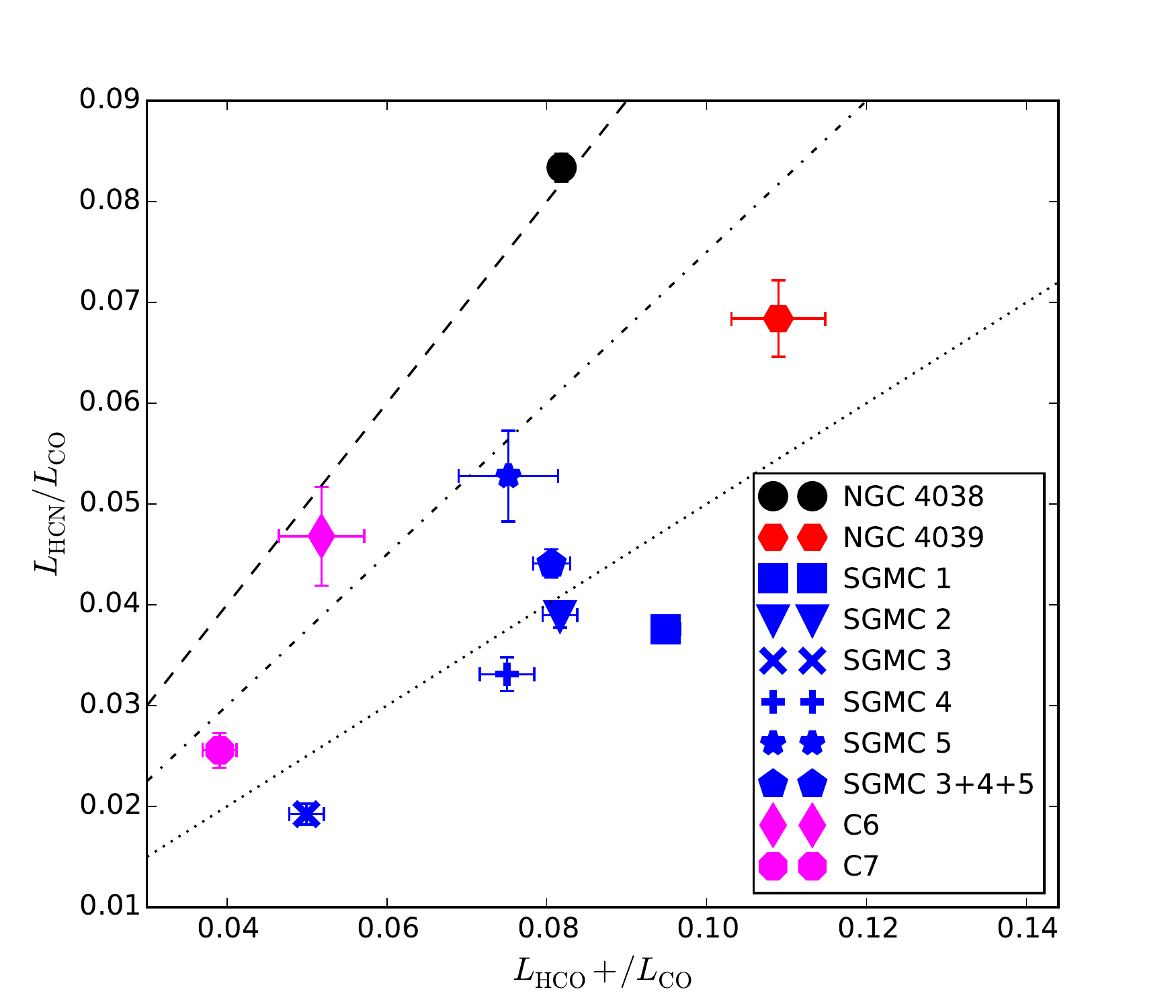}
	\caption[]{Plot of the dense gas fraction calculated at the \CO \ $J=1-0$ beam using \LNC (y-axis) and \LOC (x-axis) for the brightest regions in the Antennae. The errors bars correspond to the measurement uncertainties in the ratios, while the calibration uncertainty of these line ratios is $\sim 21 \%$. The lines correspond to lines of constant \LNO, with \LNO$= 0.5$ (dotted line), $0.75$ (dash-dotted line) and $1.0$ (dashed line). The region SGMC 3+4+5 corresponds to the region in the moment-0 map spanned by SGMC 3, 4 and 5, while the values for SGMC 3, 4 and 5 were calculated from moment maps created based on the velocity range calculated in \cite{wilson2000}}
	\label{lumRatioPlots}
\end{figure}

The ratio of the detected dense gas tracers to the total \CO \ luminosity is an indicator of the fraction of the dense molecular gas within the different regions in the Antennae. We calculate this ratio for all three dense gas tracers relative to \CO \ (Table \ref{denseRatioTable}), and plot the ratio of \LNCO \ and \LOCO \ in Figure \ref{lumRatioPlots}. 

The ratio of \LNCO \ varies by more than a factor of $4$, with the two nuclei exhibiting larger values than the other regions in the Antennae. The smallest values are seen in SGMC 3 and C7, while the remaining 6 regions have values $\sim 0.04$. The ratio of \LOCO, on the other hand, varies by a factor of $\sim 2.5$. Once again, SGMC 3 and C7, along with C6, show the smallest values of \LOCO \ ($\sim 0.04 - 0.05$), while the nucleus of NGC 4039 has the highest ratio of \LOCO \ ($\sim 0.1$). The remaining regions, including the nucleus of NGC 4038, all have a characteristic value of \LOCO \ of $\sim 0.08$.




\begin{deluxetable*}{lcccccc} 
\tablecaption{Luminosity ratios for the brightest regions in the Antennae \label{denseRatioTable}} 
\tablehead{\colhead{Region ID} & \colhead{$L_{\mathrm{HCN}}$/$L_{\mathrm{HCO}+}$} &\colhead{$L_{\mathrm{HNC}}$/$L_{\mathrm{HCO}+}$} &\colhead{$L_{\mathrm{HNC}}$/$L_{\mathrm{HCN}}$} & \colhead{$L_{\mathrm{HCN}}$/$L_{\mathrm{CO}}$} &\colhead{$L_{\mathrm{HCO}+}$/$L_{\mathrm{CO}}$} &\colhead{$L_{\mathrm{HNC}}$/$L_{\mathrm{CO}}$} 
}\startdata 
NGC 4038& $1.019 \pm 0.008$& $0.384 \pm 0.005$& $0.377 \pm 0.005$& $0.083 \pm 0.001$& $0.082 \pm 0.001$& $0.0314 \pm 0.0006$
\\NGC 4039& $0.63 \pm 0.02$& $0.160 \pm 0.008$& $0.25 \pm 0.01$& $0.068 \pm 0.004$& $0.109 \pm 0.006$& $0.017 \pm 0.001$
\\SGMC 1& $0.396 \pm 0.008$& $0.149 \pm 0.006$& $0.38 \pm 0.02$& $0.0376 \pm 0.0009$& $0.095 \pm 0.002$& $0.0142 \pm 0.0006$
\\SGMC 2& $0.48 \pm 0.01$& $0.128 \pm 0.006$& $0.27 \pm 0.01$& $0.039 \pm 0.001$& $0.082 \pm 0.002$& $0.0105 \pm 0.0006$
\\SGMC 3& $0.39 \pm 0.02$& $0.132 \pm 0.010$& $0.34 \pm 0.03$& $0.019 \pm 0.001$& $0.050 \pm 0.002$& $0.0066 \pm 0.0005$
\\SGMC 4& $0.44 \pm 0.01$& $0.149 \pm 0.008$& $0.34 \pm 0.02$& $0.033 \pm 0.002$& $0.075 \pm 0.003$& $0.0111 \pm 0.0008$
\\SGMC 5& $0.70 \pm 0.03$& $0.22 \pm 0.02$& $0.31 \pm 0.02$& $0.053 \pm 0.005$& $0.075 \pm 0.006$& $0.016 \pm 0.002$
\\SGMC 3+4+5& $0.55 \pm 0.01$& $0.173 \pm 0.006$& $0.32 \pm 0.01$& $0.044 \pm 0.001$& $0.081 \pm 0.002$& $0.0140 \pm 0.0006$
\\C6& $0.90 \pm 0.07$& $0.29 \pm 0.03$& $0.33 \pm 0.04$& $0.047 \pm 0.005$& $0.052 \pm 0.005$& $0.015 \pm 0.002$
\\C7& $0.65 \pm 0.05$& $0.24 \pm 0.03$& $0.36 \pm 0.04$& $0.026 \pm 0.002$& $0.039 \pm 0.002$& $0.009 \pm 0.001$
\enddata 
\tablecomments{The reported uncertainties are measurement uncertainties only, which are equal to the total uncertainty for the ratios which include only \HCN, \HCO \ and \HNC. The uncertainties in line ratios which include \CO \ are dominated by the calibration uncertainties. We add the calibration uncertainties from the two instruments in quadrature to obtain a total line ratio uncertainty of $21\%$. See text for additional details.} 
\end{deluxetable*}

\section{Discussion} \label{discSect}

\subsection{On the global line ratio of \LNCO}

\cite{gao2001} measured the \HCN \ $J=1-0$ emission at two locations in the Antennae with a beam FWHM $= 72 ''$: the nucleus of NGC 4038 and the overlap region, including a large portion of NGC 4039. They calculated a global value of \LNCO$= 0.02$. This is comparable to the value we calculate for SGMC 3 (\LNCO$=0.019$), the smallest value of our 9 regions. \LNCO \ ratios in five of our regions are a factor of 2 or more larger than the global value calculated by \cite{gao2001}, while the value calculated for the nucleus of NGC 4038 is a factor of 4 larger. 

There are two major contributions which could lead to such a large difference in our ratios compared to the global ratio from \cite{gao2001}: missing flux in the \CO \ interferometric observations and  \CO \ emitting gas beyond the boundaries of our defined regions. \cite{wilson2003} compared their interferometric map of \CO \ $J=1-0$ to the same region in the \cite{gao2001} single dish map and found that only $\sim 65\%$ of the flux is recovered in the interferometric map. In comparison, we determined that most of the \HCN \ emission is recovered in our map (see Section \ref{obsSect}).

The total \CO \ luminosity in our interferometric map {from the 7
  brightest regions} (with SGMC 3, 4 and 5 combined as SGMC 3+4+5) is
$L_{\mathrm{CO}} = 9.0 \times 10^8 \unit{K \ km \ s^{-1} \ pc^{2}}$,
while the total luminosity over the same area from the single dish map used in
\cite{gao2001}, scaled to our adopted distance of $22 \unit{Mpc}$, is
$L^G_{\mathrm{CO}} =  19.5 \times 10^8 \unit{K \ km \ s^{-1}
  \ pc^{2}}$ \citep{wilson2003}. Similarly, the total \HCN \ luminosity from our regions
is $L_{\mathrm{HCN}} = 5.4 \times 10^7 \unit{K \ km \ s^{-1}
  \ pc^{2}}$, while from \cite{gao2001}, $L^G_{\mathrm{HCN}} =  8
\times 10^7 \unit{K \ km \ s^{-1} \ pc^{2}}$. Based on this, $\sim 68
\%$ of the total \HCN \ emission originates from the brightest regions
of the Antennae, while only $\sim 46 \%$ of the total \CO \ emission
originates from these same regions.  

If we assume that the $\sim 32 \%$ of the \HCN \ emission found
outside of these regions is associated with \CO \ emission, we can
assume a ratio of \LNCO $\sim 0.04$ and calculate that these regions
account for another $\sim 33 \%$ of the total \CO \ emission. This
leaves $\sim 21\%$ of the \CO \ emission unaccounted for and
presumably not within relatively dense GMCs or producing significant
HCN emission. 
If we instead adopted the global value for \LNCO \ of 0.02 from
  \citet{gao2001}, then all of the \CO \ emission would originate in GMCs
  producing significant \HCN \ emission. Thus, the amount of \CO
  \ emission in the Antennae associated with diffuse molecular gas is
  $\le 20$\%.

Using the \CO \ $J=1-0$ map of M51 from the Plateau de Bure interferometer (PdBI) Arcsecond Whirlpool Survey (PAWS), \cite{pety2013} identified an extended, diffuse component of the molecular gas which account for $\sim 50\%$ of the flux, and in which \CO \ $J=1-0$ emission is subthermally excited. It is possible that a significant fraction of the $21 \%$ of the \CO \ emission that is unaccounted for in NGC 4038/39 is subthermally excited \CO \ from a diffuse, extended component such as in M51. The critical density of the \CO \ $J=1-0$ transition ($n_{cr} \sim 10^{3} \unit{cm^{-1}}$) is significantly less than that of \HCN \ ($n_{cr} \sim 10^{5} \unit{cm^{-1}}$) and so we would not expect any \HCN \ emission from an extended, diffuse component. This would lead to a suppression of the global \LNCO \ ratio when compared to smaller regions. High-resolution, flux-recovered observations of \CO \ $J=1-0$ in the Antennae using ALMA would allow us to detect the presence of this extended diffuse component.

\subsection{Dense gas fraction}

\begin{figure}[ht] 
	\centering
	\includegraphics[width = \linewidth]{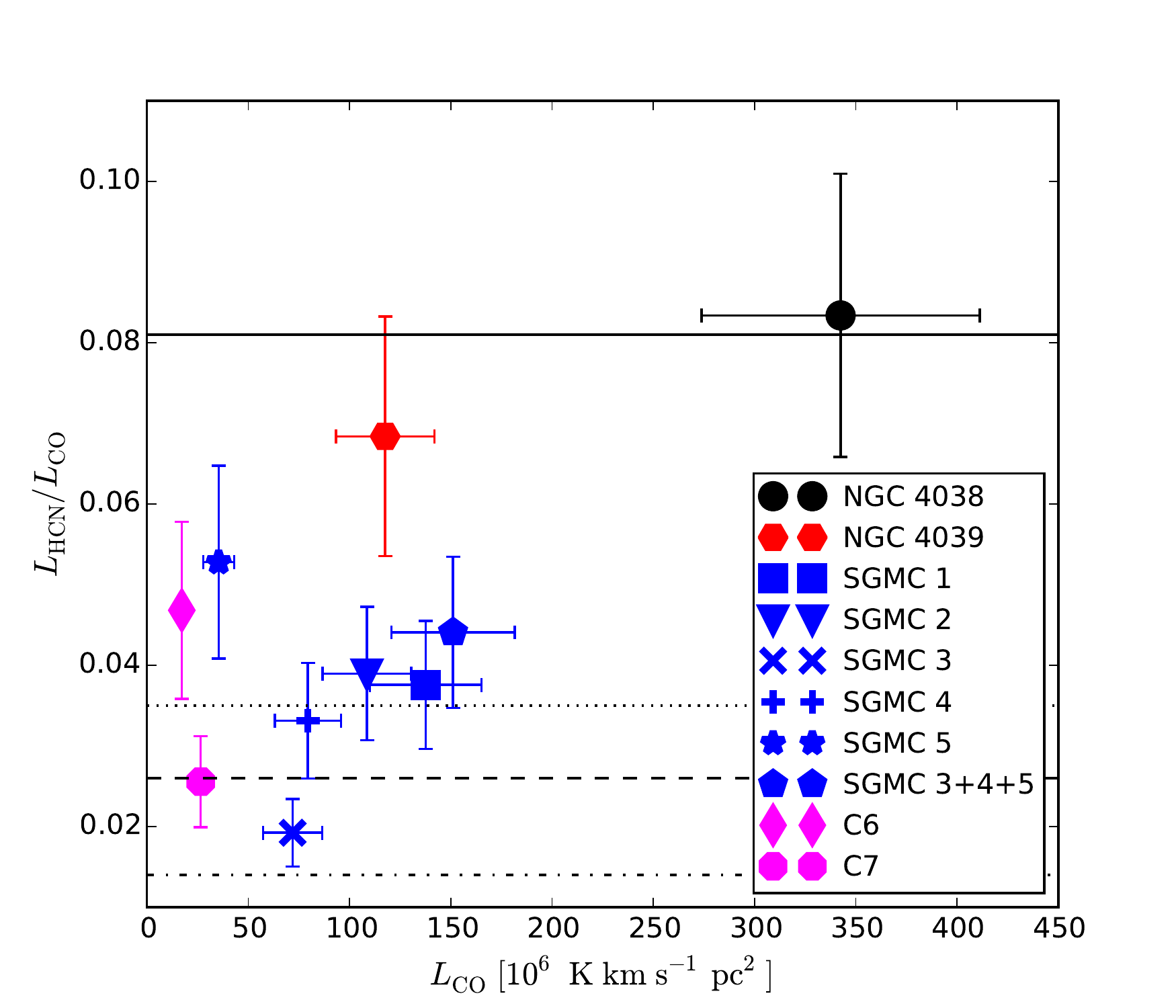}
	\caption[]{ Dense gas fraction as measured by \LNC \ (y-axis) as a function of \CO \ luminosity (x-axis). The data points are the values measured for the brightest regions in the Antennae while the error bars correspond to the measurement uncertainty and the calibration uncertainty added in quadrature. The solid and dashed horizontal lines correspond to the average values of \LNO \ as measured in the bulge and plane of the Milky Way, while the dot-dashed line corresponds to the average value measured in local GMCs \citep{helfer1997a}. The dashed line corresponds to typical values of \LNO \ in normal spiral galaxies \citep{gao2004b, graciacarpio2006}. }
	\label{HCNCO_vs_CO}
\end{figure}

We use the ratio of \LNCO \ as an indicator of the dense gas fraction throughout the Antennae, and show this ratio compared to $L_\mol{CO}$ in Figure \ref{HCNCO_vs_CO} for the 9 brightest regions in the Antennae. Typical values for normal spiral galaxies are $L_{\mol{HCN}}/L_{\mol{CO}} \sim 0.02 - 0.05$ \citep{gao2004a, gao2004b}, while this fraction can increase to $L_{\mol{HCN}}/L_{\mol{CO}} > 0.06$ in the case of some extreme star forming LIRGs and ULIRGs.

The two nuclei, NGC 4038 and NGC 4039, show a higher dense gas
fraction than the rest of the system, by roughly a factor of $\sim
2-3$. In comparison, \cite{helfer1997a} measured
$I_\mol{HCN}/I_{\mol{CO}}$, comparable to
$L_{\mol{HCN}}/L_{\mol{CO}}$, in the plane of the Milky Way and found
that the dense gas fraction was greater towards the bulge
($I_\mol{HCN}/I_{\mol{CO}} \sim 0.081$) than the rest of the plane
($I_\mol{HCN}/I_{\mol{CO}} \sim 0.026$). \cite{helfer1997b} saw a
similar increase in NGC 6946, a grand-design spiral galaxy, and NGC
1068, a Seyfert-2 galaxy with a starburst, with the ratio of
$I_\mol{HCN}/I_{\mol{CO}}$ increasing by a factor of $5-10$ towards
the bulge.  In larger surveys extending to higher infrared
luminosities, \citet{baan2008} 
report $I_\mol{HCN}/I_{\mol{CO}}$
ratios of 0.01 to 0.18 with a median spatial resolution of 5 kpc in their
sample of $\sim$75 galaxies. \citet{garcia-burillo2012} 
find ratios from
0.02 to 1 in a sample of $\sim$130 galaxies, with higher ratios more
common in galaxies with infrared luminosities above $10^{11}$
L$_\odot$. \citet{juneau2009} find ratios of 0.02 to 0.06 in
galaxies with infrared luminosities in the range $10^{10} - 10^{11}$
L$_\odot$.  Recently, \citet{usero2015} 
surveyed 62 positions in the disks of normal
spiral galaxies. They found $I_\mol{HCN}/I_{\mol{CO}}$ ratios of
0.011 to 0.065 on average scales of 1.5 kpc and found that this line
ratio
correlates with both the atomic-to-molecular gas fraction and the
stellar surface density.

\cite{helfer1997a} argue that the increased dense gas fraction in the bulge is due to an increase in the average gas pressure towards the center of the Milky Way, with $I_\mol{HCN}/I_{\mol{CO}} \propto P^{0.19}$. With this relation, a factor of 2 difference in the ratio of $I_\mol{HCN}/I_{\mol{CO}}$ would correspond to a factor of $\sim 40$ difference in the average gas pressure. This increase in gas pressure is due in part to the increased stellar density in the bulge, which increases the gravitational potential, and in turn increases the gas pressure required to support hydrostatic equilibrium. As in the Milky Way, the high ratio of $I_\mol{HCN}/I_{\mol{CO}}$ seen in the two nuclei of the Antennae could be due to an increase in the stellar density compared to the overlap region.

\cite{schirm2014} calculated the pressure using a non-LTE excitation analysis of \CO \ and \CI \ in each of NGC 4038, NGC 4039 and the overlap region in relatively large beams (FWHM $\sim 43''$, $\sim 5 \unit{kpc}$). Their cold component results suggest that the pressure in all three regions is similar ($P \sim 10^{4.5} - 10^{5.5} \unit{K \ cm^{-2}}$); however the $1\sigma$ ranges span an order of magnitude. (It is important to note that, due to the large beams, NGC 4039 contains some emission from the overlap region, and this may be reflected in the non-LTE excitation analysis.) Their warm component, traced largely by the \CO \ $J=7-6$ and $J=8-7$ transitions, shows a much higher pressure in NGC 4038 ($P \sim 10^9 \unit{K \ cm^{-2}}$) than in either NGC 4039 or the overlap region ($P \sim 10^{7.5} \unit{K \ cm^{-2}}$). 

It is possible that some, or even most, of the \HCN \ emission is associated with their warm component; however the $J=1-0$ transition is often associated with cold, dense gas. Observations of higher $J$ \HCN \ and \HNC \ transitions coupled with a non-LTE excitation analysis similar to the method used by \cite{schirm2014} would help determine whether this \HCN \ emission is associated with cold or warm dense molecular gas, while also enabling direct calculations of the pressure in the dense gas in these regions. 


We show the ratio of $I_\mol{HCN}/I_{\mol{CO}}$ in Figure
\ref{HCNHCOratioMap}, \emph{right}. In both NGC 4038 and NGC 4039,
there is a region towards the very center where the ratio of
$I_\mol{HCN}/I_{\mol{CO}}$ increases by a factor of $\sim 2$ compared
to the surrounding regions. These high ratio regions are approximately
the size of the \CO \ beam, which corresponds to $\sim 500 \unit{pc}$,
similar to the size of the bulges measured by
\cite{helfer1997b}. Interestingly, the peak value of
$I_\mol{HCN}/I_{\mol{CO}}$ is greater in NGC 4039 than in NGC 4038,
while averaged over the  whole region the dense gas fraction is greater in NGC 4038.


In the overlap region, most of the molecular gas is characteristic of
SGMC 1, where $L_\mol{HCN}/L_{\mol{CO}} \sim 0.04$. There are two
interesting peaks in the distribution of the dense gas ratio in the
overlap region, one in the south-west of SGMC 2, and another to the
south of SGMC 3+4+5.  The center of the peak in SGMC 2 is offset from
the center of pre-SSC by $\sim 3''-4''$, which corresponds to roughly
a beam. Both peaks exhibit a higher ratio of
$I_\mol{HCO+}/I_{\mol{CO}}$ and, to a lesser extent,
$I_\mol{HCN}/I_{\mol{HCO+}}$ (Figure \ref{HCN_on_Many}). No
significant emission is detected at a $3 \sigma$ level at either peak
in either \HCN \ or \HCO at the native ALMA resolution. However, both
peaks are detected in the \CO \ beam-matched maps of both dense gas
tracers at a $5 \sigma$ level. Moreover, there is no obvious \CO
\ $J=3-2$ emission detected at this location either
\citep{whitmore2014}.  It is possible that these high regions are
  produced by edge effects;  inspection of the individual
  emission maps shows that the peaks of the HCN and the CO
  emission are not precisely coincident in these areas.

\subsection{Comparison to other wavelengths}

\begin{figure*}[ht] 
\centering
$\begin{array}{ccc}
         \includegraphics[width=0.3\linewidth]{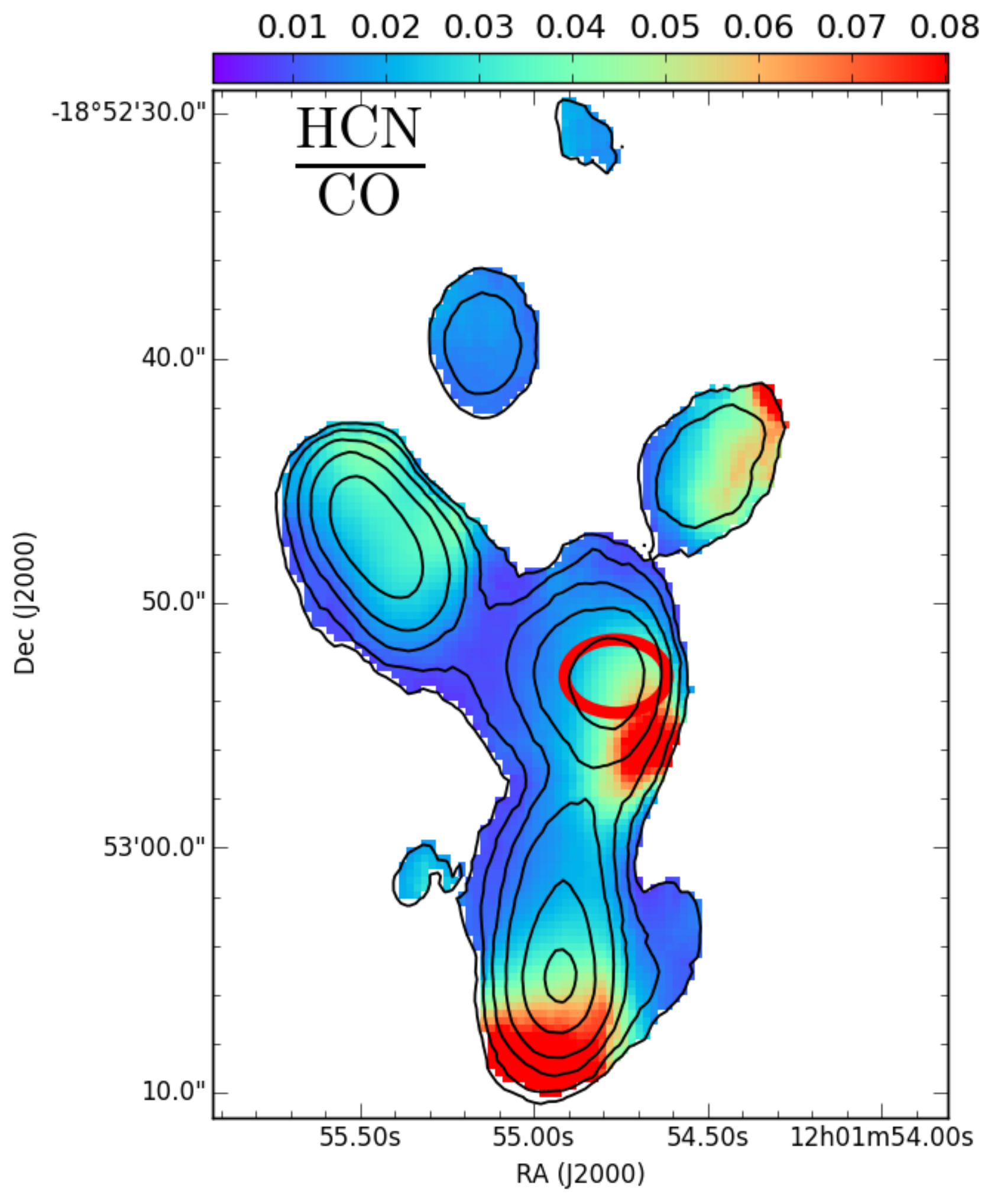} &
         \includegraphics[width=0.3\linewidth]{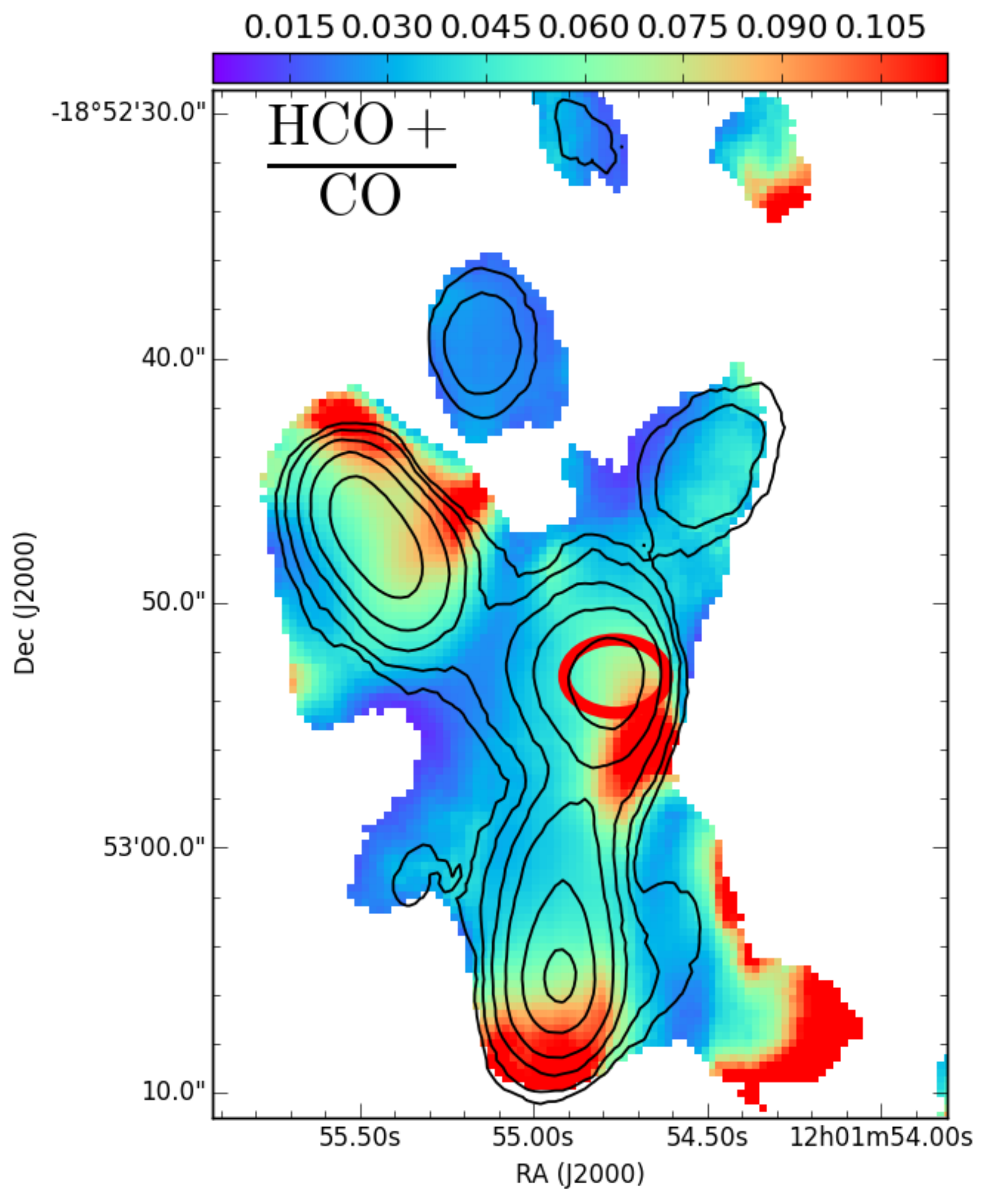} &
         \includegraphics[width=0.3\linewidth]{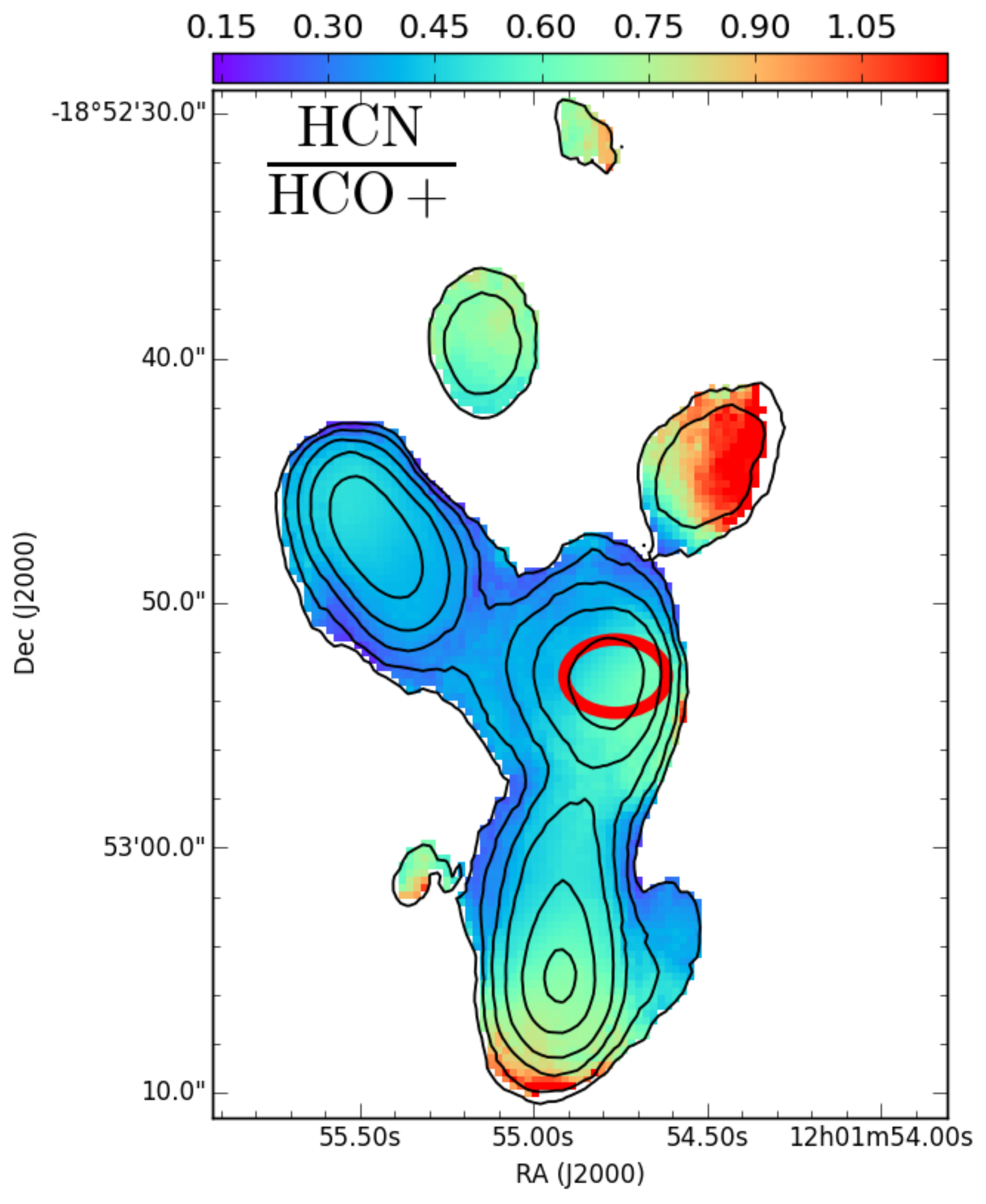} \\
	 \includegraphics[width=0.3\linewidth]{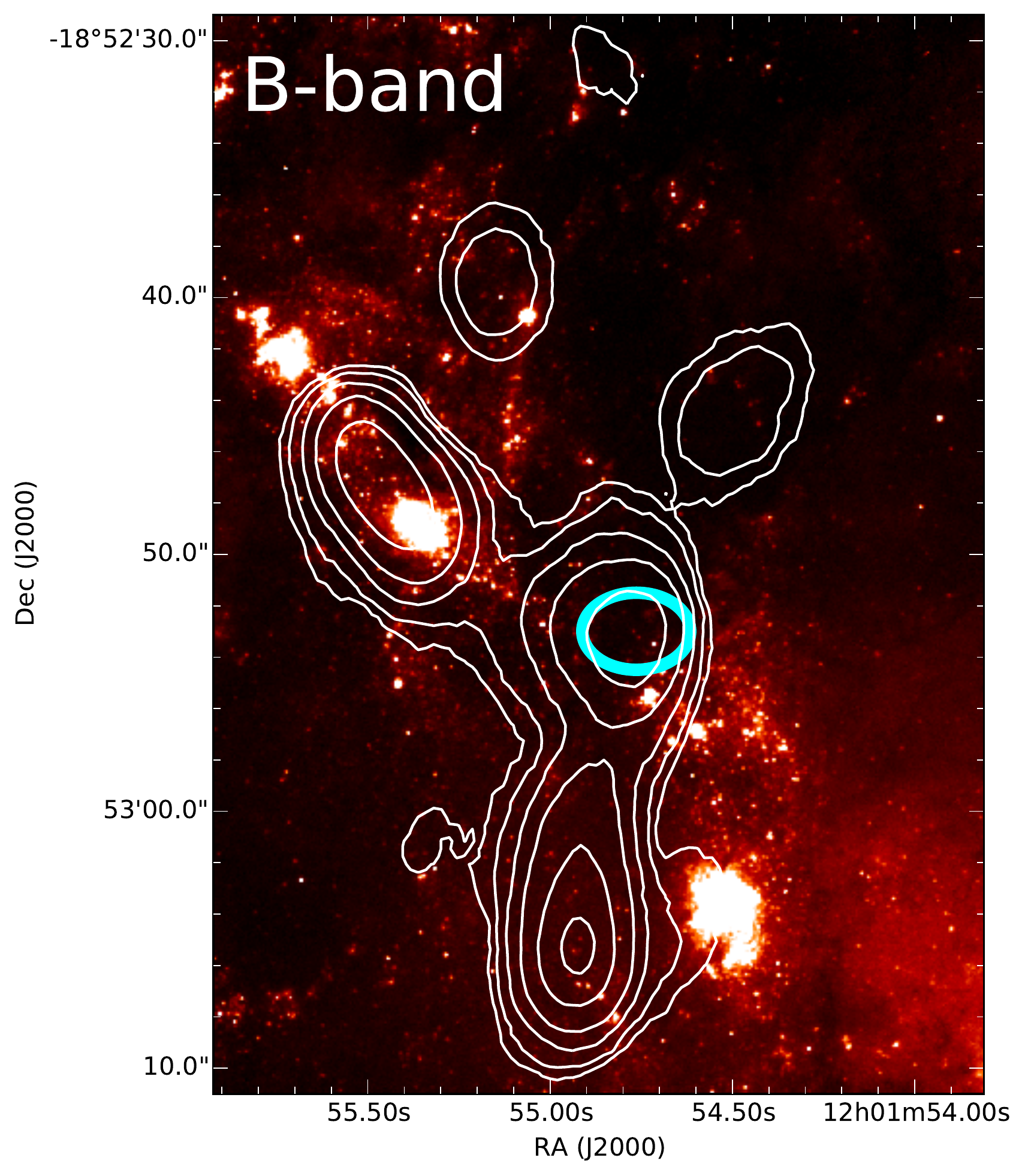} &
	 \includegraphics[width=0.3\linewidth]{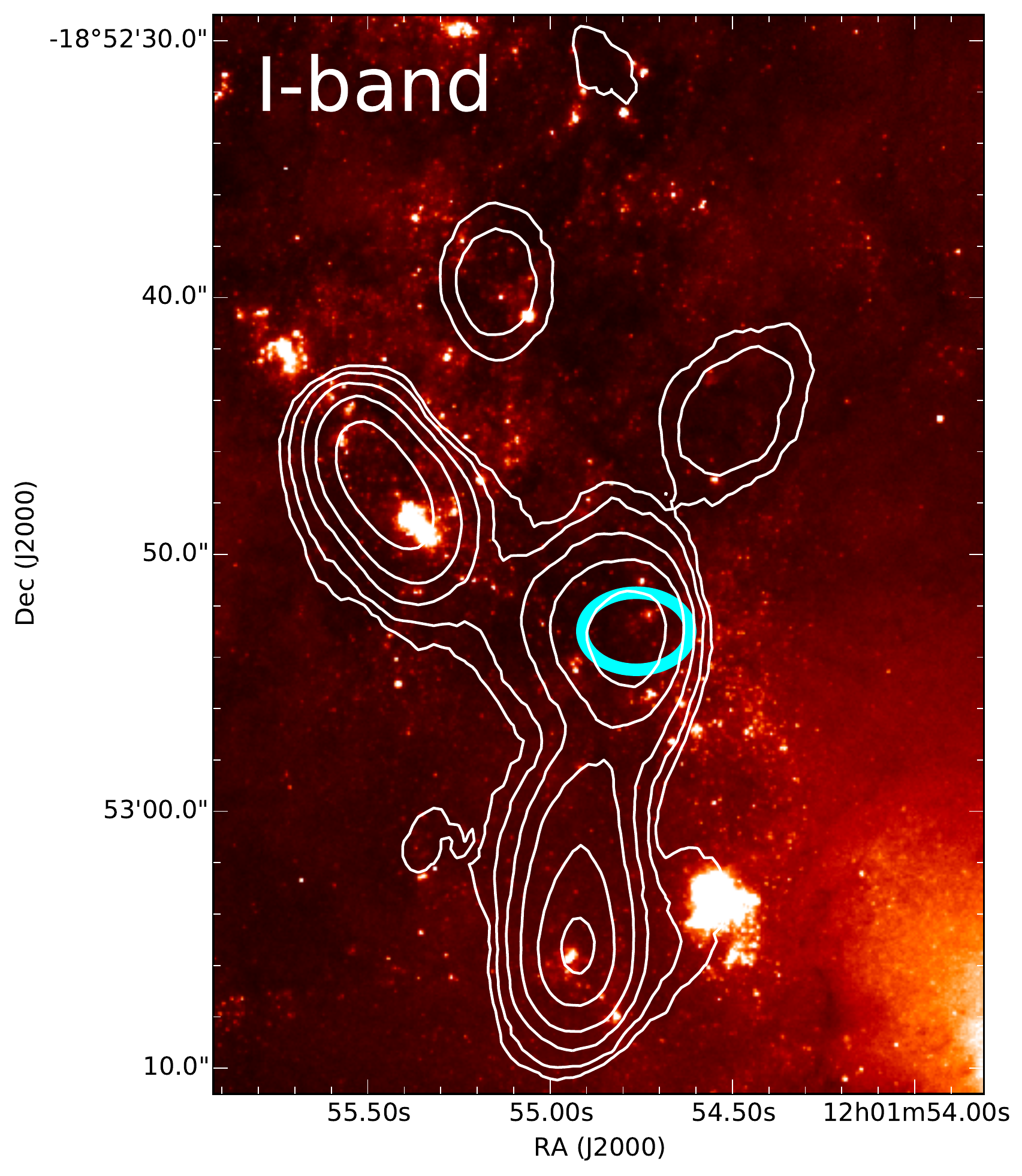} &
	 \includegraphics[width=0.3\linewidth]{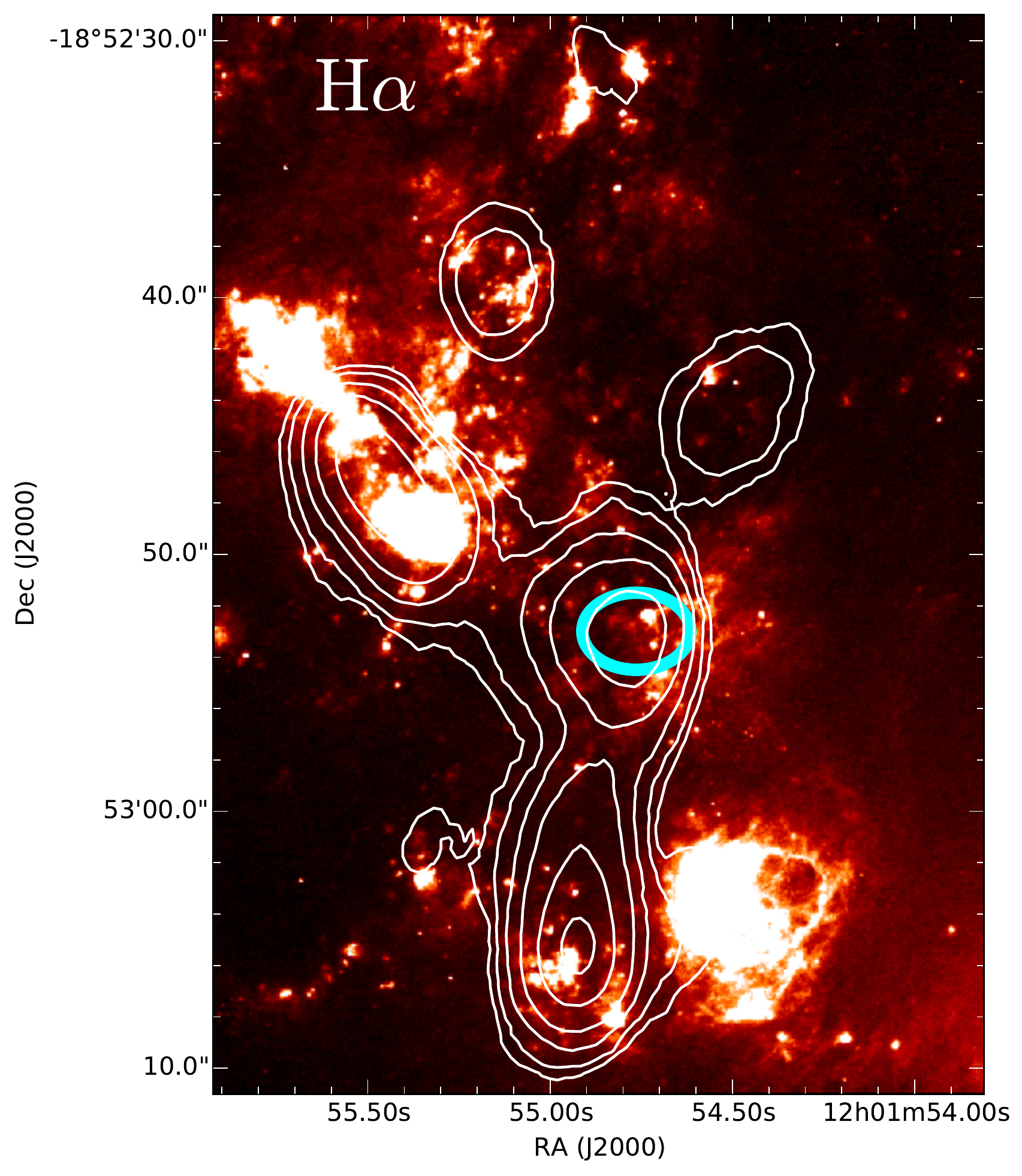} \\
	 \includegraphics[width=0.3\linewidth]{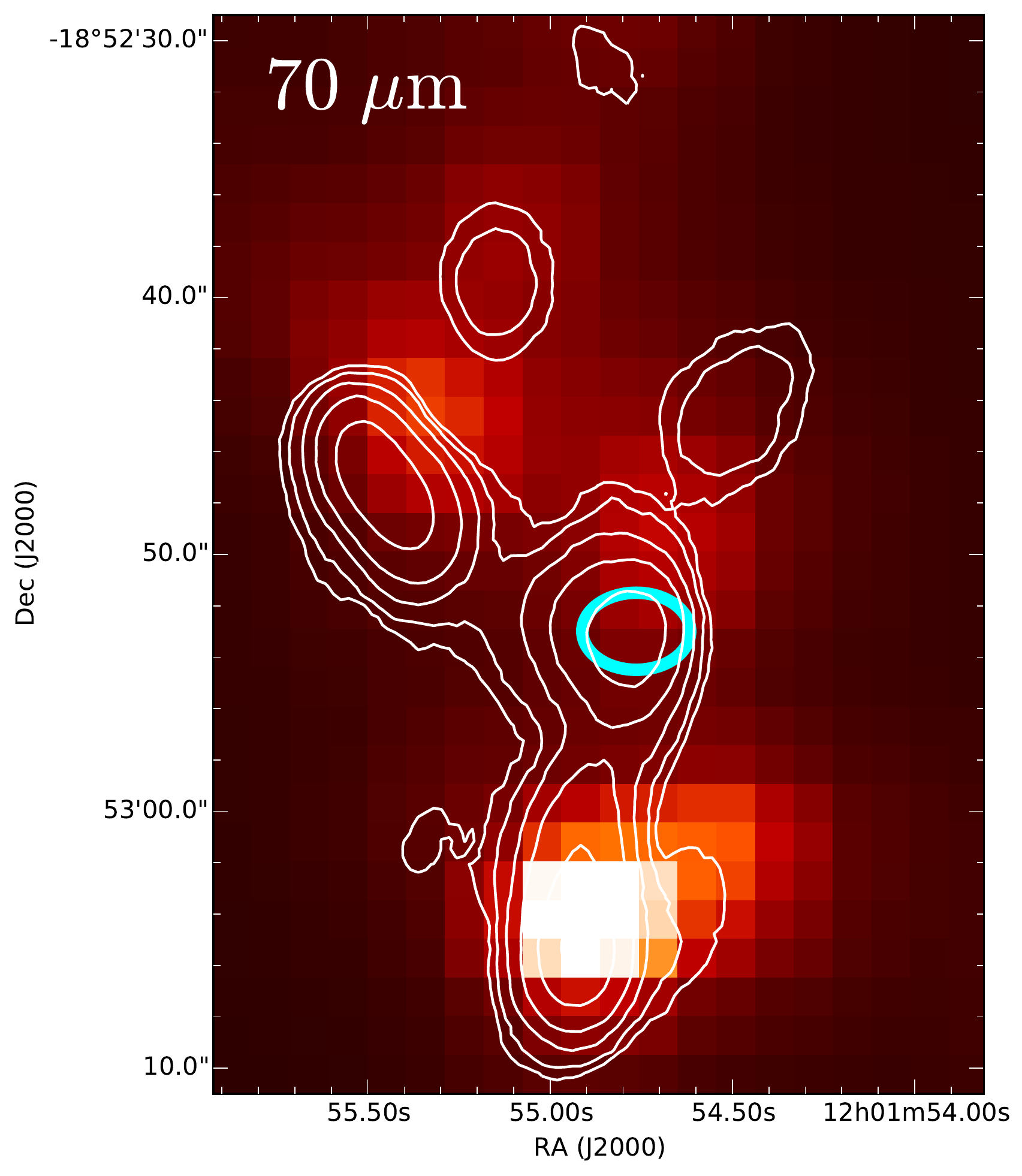} &
	 \includegraphics[width=0.3\linewidth]{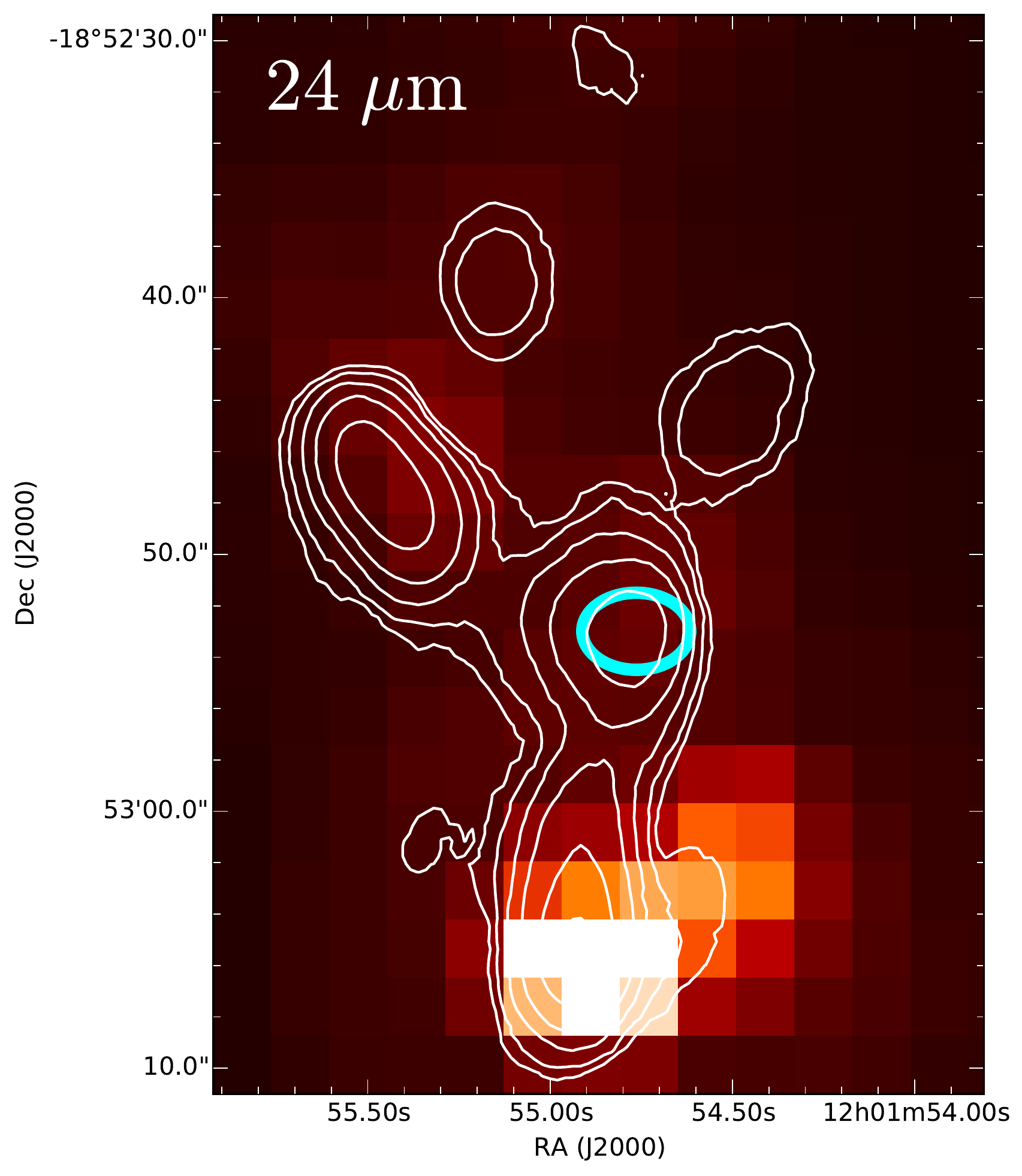} &
	 \includegraphics[width=0.3\linewidth]{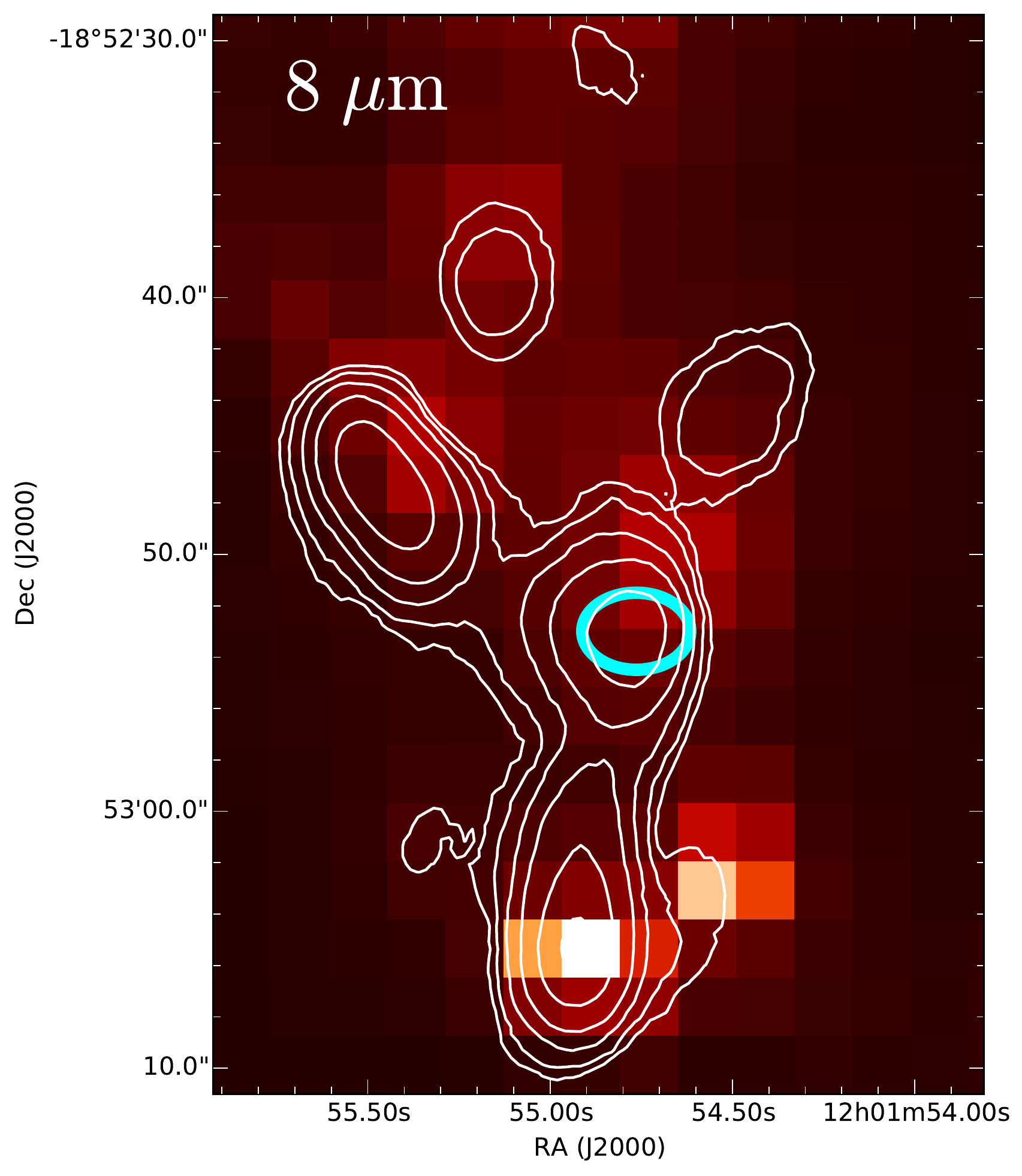}
\end{array}$
\caption[]{\HCN \ contours beam-matched to the \CO \ observations overlaid on various ratio maps and emission maps. The contours correspond to $(3, 5, 9, 15, 25, 35) \times (3.6 \times 10^{-2} \unit{Jy \ beam^{-1} \ km \ s^{-1}})$. \textbf{Top row:} Line ratio maps for \HCN/\CO \ (\emph{left}), \HCO/\CO \ (\emph{middle}) and \HCN/\HCO \ (\emph{right}). \textbf{Middle row:} Hubble Space Telescope images of the B-band (F435W, \emph{left}), I-band (F814W, \emph{middle}) and \ce{H\alpha} (F658N, \emph{right}) emission from \cite{whitmore2010}. \textbf{Bottom row:} \emph{Herschel} PACS $70 \unit{\mu m}$ emission from \cite{klaas2010} (\emph{left}), Spitzer MIPS $24 \unit{\mu m}$ emission (\emph{middle}) and Spitzer IRAC $8 \unit{\mu m}$ emission (\emph{right}), all of which trace star formation.  The location of the pre-SSC is shown by the red (\emph{top row}) and cyan (\emph{bottom row}) ellipses. None of the emission maps have any significant emission towards the high-dense fraction regions as traced by \HCN/\CO \ in the overlap region (see text for more details). }
\label{HCN_on_Many}
\end{figure*}

\cite{whitmore2010} observed the Antennae with the Hubble Space
Telescope using multiple filters, including F435W, F814W, and
F658N, which correspond to B-band, I-band and \ce{H\alpha}
emission. We compare the \HCN \ emission to each of these filters in
the middle row of Figure \ref{HCN_on_Many}.  At the two peaks
in the $I_\mol{HCO+}/I_{\mol{CO}}$ ratio discussed in the previous
section,
there are  only faint or no B-band or I-band optical counterparts.
However, there is some nearby \ce{H\alpha}
emission, at the  southern edge of the SGMC 3+4+5 peak, and near the
eastern edge of the SGMC 2 peak.  Furthermore, there is no $70
\unit{\mu m}$ \citep{klaas2010}, $24 \unit{\mu m}$ \citep{zhang2010}
or $8 \unit{\mu m}$ \citep{wang2004} emission associated with the line
ratio peak
near SGMC-2, although there is bright emission in these bands towards
SGMC 3+4+5 that could have some relation to the southern peak in the
line ratio (Figure \ref{HCN_on_Many}). 

The lack of emission in the B- and I-band images
is likely due to high dust extinction in the
overlap region. The $70 \unit{\mu m}$ \citep{calzetti2010}, $24
\unit{\mu m}$ and $8 \unit{\mu m}$ \citep{wu2005b} fluxes all
correlate strongly with star formation, and so we attribute the
deficiency in their emission to a lack of recent star formation within
one or possibly both of the line ratio peaks in the overlap
region. As such, these 
high-dense gas fraction regions may be the result of turbulent
motion due to the ongoing merger, and may indicate sites of future
star formation within the overlap region. 

\subsection{Mechanical heating and photon dominated regions}

\HCN \ and \HNC \ are isotopomers with similar excitation energies. The line ratio of \LCN \ is driven in large part by their relative abundance, in which the exchange reaction  \ce{H + HNC <-> H + HCN} is an important factor \citep{schilke1992, talbi1996}. At temperatures $<100 \unit{K}$, the rate coefficients of both the forward and reverse reactions are small, and this reaction is not important \citep{talbi1996}. At temperatures of a few hundred $\mol{K}$, the energies are sufficient to exceed the activation energy of the \ce{H + HNC} reaction, leading to a higher relative abundance of \HCN \ to \HNC \ and a lower ratio of \LCN. 

It has been suggested that most molecular gas is in the form of PDRs (e.g. see the PDR model comparison study by \citealt{roellig2007}). If mechanical heating is not included in the model, heating occurs only at the surface of the PDR, where the photon energies are high enough to liberate elections from the surfaces of dust grains. Recent PDR models have investigated the effects of mechanical heating on the molecular gas within PDRs \citep{meijerink2011, kazandjian2012, kazandjian2015}, in particular on the predicted atomic and molecular line ratios. They find that mechanical heating begins to have a measurable effect on the chemistry of the PDR, and consequently the line ratios, with contributions of as little as $\sim 1\%$ to the total heating \citep{kazandjian2012}. 

In the Antennae, the ratio of \LCN \ varies by less than a factor of $1.5$ across the brightest regions, ranging from $\sim 0.25 - 0.38$ (Table \ref{denseRatioTable} and Figure \ref{denseRatPlot}). We compare our measured ratio to the PDR models with mechanical heating from  \cite{kazandjian2015}. In their models, the mechanical heating is parameterized as $\alpha$, which is equal to the total amount of mechanical heating relative to the PDR surface heating. For values of $\alpha<0.05$, their reference models report a value of \LCN $ \ge 1$. Our measured line ratios are consistent with at least some ($>5\%$) contribution of mechanical heating towards the total heating budget in all the bright regions in the Antennae.

In comparison, \cite{schirm2014} modelled various line ratios of \CO
\ from $J=1-0$ to $J=8-7$ along with the FIR luminosity using PDR
models which did not include mechanical heating. In all three region,
they found that line ratios of the low-J \CO \ transitions ($J=1-0$ to
$J=3-2$) are consistent with a ``cold'' PDR with a field strength of
$G_0 \sim 100$ and a molecular gas density of $n(\mol{H_2}) \sim 10^3
- 10^4 \unit{cm^{-3}}$. Line ratios of the high-J \CO \ transitions
($J=6-5$ to $J=8-7$) are consistent with a ``warm'' PDR with a field
strength of $G_0 \sim 1000$ and a molecular gas density of
$n(\mol{H_2}) \sim 10^4 - 10^5 \unit{cm^{-3}}$, while the warm PDR
accounts for $\sim 1\%$ of the total molecular gas mass. Their warm
PDR solutions are similar to model M3 from \cite{kazandjian2015}, for
which \LCN \ $\sim 1$ for all amounts of mechanical heating, while the field strength of their cold PDR models lies outside of the range of $G_0$ modelled by \cite{kazandjian2015}.  Future PDR modelling within the Antennae should include mechanical heating as it has to have a fundamental effect on the chemistry within these regions. 

The higher ratio of \LCN \ seen in the nucleus of NGC 4038 and SGMC 1 indicates that the relative contribution of mechanical heating is lower within these regions. Conversely, the lower ratio seen in the nucleus of NGC 4039 and SGMC 2 indicates a higher relative contribution of mechanical heating. An increase (decrease) in the relative contribution of mechanical heating could either be due to an increase (decrease) in the total mechanical heating, or due to a decrease (increase) in the total PDR surface heating. All four regions show similar star formation efficiencies (SFEs), ranging from $2.19 \unit{L_\odot M_\odot^{-1}}$ for NGC 4038, to $3.33 \unit{L_\odot M_\odot^{-1}}$ for SGMC 2 \citep{klaas2010}. As PDR surface heating is tied to the background FUV field strength, which is emitted from young, massive stars, the similar SFEs within these four regions likely indicate variations in the amount of mechanical heating as opposed to the PDR surface heating. This, in turn, agrees with the results from \cite{schirm2014} who found similar values for $G_0$ and $n(\mol{H_2})$ in NGC 4038, NGC 4039 and the overlap region. 

Finally, our ratios of \LCN \ are consistent with models from \cite{kazandjian2015} with $\alpha > 0.1$ (while also consistent with $0.1 > \alpha > 0.05$). The molecular gas temperature of the models with $\alpha > 0.1$ is high ($T_{kin} > 100 \unit{K}$), which would indicate that the \HCN \ and \HNC \ emission originates from warm, dense molecular gas similar to the warm component found by \cite{schirm2014}. However, the $J=1-0$ transition of both these molecules is typically assumed to be associated with cold, dense, star forming molecular gas. It could be that this cold, dense gas is quickly heated by the ongoing star formation within these systems, via supernovae and stellar winds, or that the turbulent motion due to the ongoing merger heats this dense gas. A multi-transitional non-LTE excitation analysis can be used to assess whether the \HCN \ and \HNC \ $J=1-0$ emission is from cold or warm dense gas. 


\subsection{Cosmic rays and the abundances of \HCN \ and \HCO}

Differences in the ratio of \HCN \ and \HCO \ of over a factor of $2$
are seen in both the total global line ratio in the form of \LNO
\ (Table \ref{denseRatioTable}) and in the distribution of
$I_{\mol{HCN}}/I_{\mol{HCO+}}$ (Figure \ref{HCNHCOratioMap}). In
particular, the nucleus of NGC 4038 exhibits a ratio $\sim 1$, while
in the SGMCs, the ratio is typically $\sim 0.4-0.6$. These differences
in the line ratio suggest either different excitation conditions for
the two molecules (e.g. see \citealt{juneau2009}) or changes in the
relative abundances. To look for varying excitation conditions,
additional transitions are required of both molecules. Thus, we
discuss the implications of varying abundances of \HCN \ and \HCO \ in
the context of the measured line ratios.  

\HCO \ is an ion and is easily destroyed through recombination in the
presence of free electrons. Free electrons can also combine with
\ce{HCNH+} to form \HCN \ \citep{lintott2006, juneau2009},
simultaneously enhancing the abundance of \HCN \ while suppressing the
abundance of \HCO.  These free electrons are generated in the
formation of the $\mol{H}_3^+$ ion, which occurs via $2\mol{H_2} +
\zeta \rightarrow \mol{H}_3^+ + \mol{H} + e^-$ \citep{mccall2003}. \cite{papadopoulos2007} argue that deep within molecular clouds, where the dense gas is found, this reaction with cosmic rays is the primary source of free electrons; however $\mol{H}_3^+$ is also important in the formation of \HCO. Thus, interpreting variations in the line ratio of \HCN \ and \HCO \ is not as straightforward as, say, \HNC \ and \HCN. 

\cite{meijerink2011} investigated the effects of cosmic rays on the
abundances of various molecules, atoms and ions within PDRs. They
modelled a dense ($n(\mol{H}_2) = 10^{5.5} \unit{cm^{-3}}$) PDR region
with a strong background UV field ($10^5 G_0$), along with an
intermediate density ($n(\mol{H}_2) = 10^{3} \unit{cm^{-3}}$) PDR with
a PDR field strength of $G_0 = 10^3$. They varied the cosmic ray rate from $5\times 10^{-17} \unit{s^{-1}}$ to $5\times 10^{-13} \unit{s^{-1}}$. In the dense PDR, they found that the abundances of both \HCN \ and \HNC \ are insensitive to changes in the cosmic ray rates. The abundance of \HCO, however, increases with increasing cosmic ray rates, except at very high cosmic ray rates. In the intermediate density PDR case, the \HCO, \HCN \ and \HNC \ abundances all decrease with increasing cosmic ray rate. 

Cosmic rays are assumed to originate largely from supernova remnants
\citep{schulz2007}. \cite{neff2000} measured the supernova rate
($\nu_{SN}$) across the Antennae using nonthermal radio sources which
trace compact supvernova, and found a global rate of $\nu_{SN} \sim
0.2 - 0.3 \unit{yr^{-1}}$. \cite{schirm2014} compared the location of
these supernova to their $\sim 43''$ beams for each of NGC 4038, NGC
4039 and the overlap region and determined that $\sim 66\%$ of the
supernova originate from the overlap region, $14 \%$ from the nucleus
of NGC 4038 and $\sim 6\%$ from the nucleus of NGC 4039. 
In our new ALMA maps, the \HCO \ emission in the overlap region
  covers roughly twice 
  the area of \HCO \ emission in either of the two nuclei. 
\citet{neff2000} identified 115 sources in their 6-cm map, of which
they estimate 60\% to be non-thermal sources; there are 18 source
bright enough for their spectral index to be measured with sufficient
accuracy to show they are non-thermal. Comparing to their catalogs,
the overlap region contains 36 sources, of which 6 are confirmed to be
non-thermal, NGC 4038 has 9 (4) sources, and NGC 4039 has 5 (1)
sources. These source counts suggest that the surface density of
supernova in the overlap region is 0.7-2 times that of NGC 4038, while
the surface density in NGC 4039 is 0.3-0.6 times that of NGC 4038,
with the lower values corresponding to sources with confirmed
non-thermal spectral indices.


Unfortunately, there is no clear trend between the \LNO ratio and the
supernova surface density among these three sources. If the overlap
region has a higher supernova surface density than NGC 4038, this
could produce an increase in the local cosmic ray rate. This could in turn
explain the decreased ratio of \LNO \ found in the overlap region. 
However, NGC
4039 has a lower supernova surface density than either of the
  other two regions
and yet exhibits an \LNO \ ratio that is more similar to that of
  the overlap region. An active galactic
nucleus (AGN) can also be the source of cosmic rays \citep{auger2007};
however multi-wavelength studies of the Antennae suggest that there is
no AGN at the center of NGC 4039 \citep{neff2000, brandl2009,
  ueda2012}. \cite{ueda2012} suggest that the high line ratio of \CO
\ $J=3-2$/$J=1-0$ seen in NGC 4039 is not due to star formation
activity, but could possibly be due to a hidden AGN. If that is the
case, the lower value of \LNO \ seen in NGC 4039 could perhaps be
explained by the presence of such a hidden AGN.

\section{Summary and conclusions} \label{conc}

In this paper, we present high-resolution observations of the dense gas tracers \HCN, \HCO, and \HNC \ $J=1-0$ transitions in the Antennae using ALMA.  
These observations are beam matched and compared to previously obtained lower resolution \CO \ $J=1-0$ observations by \cite{wilson2000, wilson2003}. We isolate the emission from the nucleus of NGC 4038 and NGC 4039, and from the 5 SGMCs in the overlap region. We also identify two other bright regions, clouds C6 and C7, located to the north of the overlap region.

\begin{enumerate}

	\item We compare our interferometric observations of \HCN \ and \CO \ $J=1-0$ to single-dish observations of the same transitions by \cite{gao2001}. We find that $\sim 68\%$ of the total \HCN \ flux from the \cite{gao2001} observations is located within the nuclei, SGMCs and clouds C6 and C7, while  $\sim 46 \%$ of the \CO \ emission is from these same regions. Furthermore, assuming a line ratio of \LNCO$\sim 0.04$, we find that there may be up to $20 \%$ of the \CO \ emission not associated with any \HCN \ emission. We suggest that this \CO \ is subthermally excited and the emission originates from relatively diffuse molecular gas, similar to that seen in M51. 
	

\item The dense gas fraction as measured by \LNCO \ is higher in the two nuclei ($0.083$ and $0.068$ respectively) than in any other region of the Antennae ($<0.053$). Furthermore, the line ratio peaks in the centre of the two nuclei. This increase is consistent with what is seen within the bulges of nearby spiral galaxies, where the stellar potential is larger. We attribute this increase in the dense gas fraction to an increase in the pressure within the two nuclei due to the higher stellar potential in the bulge region.  

\item The ratio of \LCN \ is a tracer of mechanical heating within PDRs. We find that this ratio varies by less than a factor of 1.5 across our defined regions in the Antennae, ranging from $0.25 - 0.38$. By comparing these values to PDR models which include mechanical heating, we find that mechanical heating must be at least $5\%$ of the PDR surface heating in the Antennae.  In these PDR models, a minimum contribution of $10\%$ of mechanical heating relative to PDR surface heating indicates temperatures $>100\unit{K}$, which is consistent with the values we measure. This may indicate that both \HCN \ and \HNC \ $J=1-0$ are tracers of warm, dense molecular gas in the Antennae as opposed to cold, dense molecular gas. A multi-transitional non-LTE excitation analysis would be useful to determine the temperature of this gas. 

\item The \LNO ratio peaks in the nucleus of NGC 4038 (\LNO $\sim
  1.0$), while it is approximately a factor of 2 smaller in the
  overlap region. This difference  may be due to a increase in
  the abundance of \HCO \ in the overlap region due to an increase in
  the cosmic ray rate from an increased supernova rate. We also find a
  difference in the ratio between NGC 4038 and NGC 4039 ($\sim
  0.6$). This difference seems unlikely to be due to an increased
  supernova rate; a
  hidden AGN in NGC 4039 could be an additional source of cosmic
  rays. However, few studies have shown any evidence of a
  hidden AGN, and so its existence remains speculative.

\end{enumerate}


%
%
%
%
%
%
%
%

We thank the referee for comments which substantially improved
  the discussion and analysis. This paper makes use of the following
  ALMA data: ADS/JAO.ALMA\#2012.1.00185.S.
ALMA is a partnership of ESO (representing its member states), NSF
(USA) and NINS (Japan), together with NRC (Canada), NSC and ASIAA
(Taiwan), and KASI (Republic of Korea), in cooperation with the
Republic of Chile. The Joint ALMA Observatory is operated by ESO,
AUI/NRAO and NAOJ. The National Radio Astronomy Observatory is a
facility of the National Science Foundation operated under cooperative
agreement by Associated Universities, Inc. The research of C.D.W. is
supported by NSERC Canada.  
This research made use of the python plotting package matplotlib
\citep{hunter2007}. This research made use of APLpy, an open-source
plotting package for Python hosted at http://aplpy.github.com.

\bibliographystyle{mn2e}
\bibliography{NGC4038ALMA_Schirm_rev}

\end{document}